\documentclass[twocolumn,showpacs]{revtex4}

\usepackage[latin1]{inputenc}
\usepackage{graphicx}%
\usepackage{dcolumn}
\usepackage{amsmath}

\makeatletter
\def\btt#1{\texttt{\@backslashchar#1}}%
\DeclareRobustCommand\bblash{\btt{\@backslashchar}}%
\makeatother

\begin{document}


\title{Criterion for traffic phases in single vehicle data and empirical test of a microscopic three-phase traffic theory}

\author{Boris S. Kerner $^1$,  Sergey L. Klenov $^2$, and Andreas Hiller $^3$}


\affiliation{$^1$, $^3$
DaimlerChrysler AG, REI/VF, HPC:  G021, 71059 Sindelfingen, Germany 
}

\affiliation{$^2$
Moscow Institute of Physics and Technology, Department of Physics, 141700 Dolgoprudny,
Moscow Region, Russia
}



\begin{abstract}
A microscopic criterion 
for distinguishing   synchronized flow and   wide moving jam phases 
 in single vehicle data  measured at a single freeway location is presented. Empirical local congested traffic states
  in single vehicle data measured on different days are classified into synchronized flow
  states and states consisting of synchronized flow and wide moving jam(s).
  Then empirical microscopic   characteristics 
   for these different local congested traffic states are studied.
   Using these    characteristics and empirical
      spatiotemporal  macroscopic traffic phenomena,
     an empirical test of a microscopic three-phase traffic flow theory  is performed. 
  In accordance with real traffic, a model of an $\lq\lq$open" road
is applied; traffic demand at   model boundaries
is taken from empirical data. Spatiotemporal 
congested model patterns emerge,  develop, and dissolve due to self-organization effects.  
In accordance with the microscopic criterion for the traffic phases,
  the synchronized flow and   wide moving jam phases are found in single vehicle model data
  associated with different  locations within the spatiotemporal congested model patterns. 
  Simulations show that the microscopic criterion and macroscopic spatiotemporal objective criteria
  lead to the same identification of the synchronized flow and   wide moving jam phases in congested traffic.
It is found that  microscopic three-phase traffic models can explain   both microscopic and macroscopic
empirical congested pattern features.   
It is obtained that microscopic frequency distributions for vehicle speed difference
as well as fundamental diagrams and speed correlation functions can depend on the spatial co-ordinate considerably. 
It turns out 
that  microscopic optimal velocity (OV)  functions and  time headway distributions are not necessarily   qualitatively different,
even if   local    congested traffic states are qualitatively different. 
The reason for this is that important {\it spatiotemporal} features of congested traffic patterns are   {\it lost}  in these as
well as in many other 
macroscopic and microscopic traffic characteristics, which are widely used as the empirical basis for a test of traffic flow models,
specifically, cellular automata traffic flow models. 
\end{abstract}

\maketitle

\section{Introduction}
\label{Intr}

Freeway traffic is a complex dynamic process, which unfolds in space and time. Specifically, 
in   initial free flow complex spatiotemporal congested patterns are observed. Therefore, for an adequate comparison with
reality
a model should exhibit features of  the onset of congestion and of empirical {\it spatiotemporal} congested traffic patterns. To observe these patterns,
measurements of traffic variables (e.g., flow rate and vehicle speed) as functions of time 
 at many different freeway locations should  simultaneously be made on a long enough freeway section. From spatiotemporal analysis of 
such data measured over many days and years on various freeways in different countries,  it has been found that
there are two different phases in congested traffic, synchronized flow and wide moving jam
(see references in the book~\cite{KernerBook}). Thus, 
there are three-traffic phases: 1. Free flow. 2. Synchronized flow. 3. Wide moving jam. 

The fundamental difference between
synchronized flow and wide moving jam is determined by the following macroscopic spatiotemporal objective (empirical)   
  criteria~\cite{KernerBook}. 
The wide moving jam exhibits the
\emph{characteristic, i.e., unique, and coherent feature}  to maintain the mean velocity of the downstream jam front, 
 even  when the jam propagates through any other
traffic states or  freeway bottlenecks.   
In contrast, 
 synchronized flow  does not exhibit this characteristic feature, in particular, the downstream front of synchronized flow
is often  \emph{fixed} at a freeway bottleneck.

Congested traffic occurs mostly  at freeway bottlenecks. Just as defects and impurities are important for 
phase transitions in physical systems, so are bottlenecks in traffic flow. In empirical observations, the following fundamental spatiotemporal
features of phase transitions and congested patterns have been found~\cite{KernerBook}: 

(1) The onset of congestion at a bottleneck is associated with
a local  phase transition from free flow to synchronized flow. This F$\rightarrow$S transition
exhibits probabilistic nature.
In particular, the F$\rightarrow$S transition can be induced by a short-time disturbance that plays the role of a critical nuclei
for the phase transition. 

(2) There can be   spontaneous and induced F$\rightarrow$S transitions at a freeway bottleneck.

(3) Wide moving jams can emerge spontaneously only in synchronized flow, i.e., due to F$\rightarrow$S$\rightarrow$J transitions. 

(4) There are two main types of congested patterns at an isolated bottleneck: A synchronized flow pattern (SP) and a general pattern (GP). 
An SP consists of synchronized flow only, i.e., no wide moving jams emerge in the SP. An GP is a congested pattern, which consists of synchronized flow
upstream of the bottleneck and wide moving jams that emerge spontaneously in this synchronized flow. 

To explain the complex dynamic process of freeway traffic, a huge number of traffic flow models have been introduced.
The last few years have seen a rapid development of traffic flow physics
in relation to new modeling approaches (see the reviews~\cite{Gartner,Wolf,Sch,Helbing2001,Nagatani_R,Nagel2003A,MahnkeRev}, the book~\cite{KernerBook}, and the conference proceedings~\cite{Lesort,Ceder,Taylor,SW1,SW2,SW3,SW4,SW5}).

Past traffic flow theories and models reviewed in~\cite{Gartner,Wolf,Sch,Helbing2001,Nagatani_R,Nagel2003A}
cannot explain and predict the spatiotemporal features of traffic mentioned in item (1)--(4) above. The only one exception is the characteristic 
features of wide moving jam propagation that can be shown in some of these models~\cite{KernerBook}.
To explain  empirical results of item (1)--(4), Kerner introduced three-phase traffic theory~\cite{Kerner1998B,Kerner2002B}.
The hypotheses of this theory are the basis of a microscopic three-phase traffic theory~\cite{KKl,KKW,KKl2003A,KKl2004AA}. 
Recently, some new microscopic models in the context of three-phase traffic
theory
have been suggested~\cite{Davis2003B,Lee_Sch2004A,Jiang2004A}.

Microscopic three-phase traffic theory of Ref.~\cite{KKl,KKW,KKl2003A,KKl2004AA} can explain macroscopic spatiotemporal
congested pattern phenomena~\cite{KernerBook}.
Macroscopic traffic flow characteristics can  depend strongly on the pattern type    and on the freeway location within the
pattern. Thus, we can expect that this conclusion is also valid for   single vehicle (i.e., microscopic) traffic flow characteristics.
However, single vehicle data measured at  many different freeway locations, which can be sufficient for 
  a reliable  spatiotemporal analysis of the whole congested pattern structure, are not available up to now.

In this paper, a   microscopic empirical criterion 
for the synchronized flow and  wide moving jam phases in congested traffic is found.
This criterion enables us to distinguish the phases in empirical   single vehicle data
measured at a {\it single} freeway location, i.e., even without a spatiotemporal
congested pattern analysis required for application of the macroscopic  spatiotemporal objective criteria for the traffic phases.
Based on the associated empirical single vehicle data analysis,   
an empirical test of a microscopic three-phase traffic theory of Ref.~\cite{KKl,KKW,KKl2003A,KKl2004AA} is performed.

The article is organized as follows. 
Empirical findings   are  considered in Sect.~\ref{EmpTest_Sect}.
Firstly, a microscopic empirical criterion for the phases in congested traffic
is suggested and used for an empirical single vehicle data analysis (Sect.~\ref{Interruption}). Based on this analysis,
empirical microscopic    characteristics of local congested  states are found
(Sect.~\ref{Single_Dep_Sect}). Finally, some of the fundamental empirical  macroscopic 
  spatiotemporal pattern features are briefly discussed (Sect.~\ref{STEmpTest_Sect}).   
  In Sect.~\ref{MST_Sect}, microscopic and macroscopic congested model patterns and their characteristics 
  are compared with empirical
results. Firstly, model   spatiotemporal congested pattern evolution
under time-dependence of traffic demand taken from empirical results is studied
and compared with the empirical macroscopic spatiotemporal patterns of Sect.~\ref{STEmpTest_Sect}. 
Then these spatiotemporal congested patterns are used to find
model microscopic characteristics  from  single vehicle data associated with different locations within
the patterns. Finally, a proof of the microscopic criterion for the phases in congested traffic as well as
 a comparison of empirical and microscopic traffic model characteristics are presented.

\section{Empirical Results}
\label{EmpTest_Sect}

\subsection{Congested States in Single Vehicle Data}
\label{Con_states}

Single vehicle characteristics are usually obtained either in driver experiments or through the use of 
detectors (e.g.,~\cite{Neubert,Knospe2002,Knospe2004,Cowan,Koshi,Luttinen,Bovy,Tilch2000,Banks,Gurusinghe2003A}).
In the latter case, data from a single freeway location or aggregated data measured at different locations 
are used~\cite{Aggregated}.
For empirical tests of traffic flow models time headway distributions (probability density for different time headways)
and optimal velocity (OV) functions (the mean speed as a function of space gap between vehicles at a given density)
are often used (e.g.,~\cite{Knospe2004}).

\begin{figure*}
\begin{center}
\includegraphics[width=12 cm]{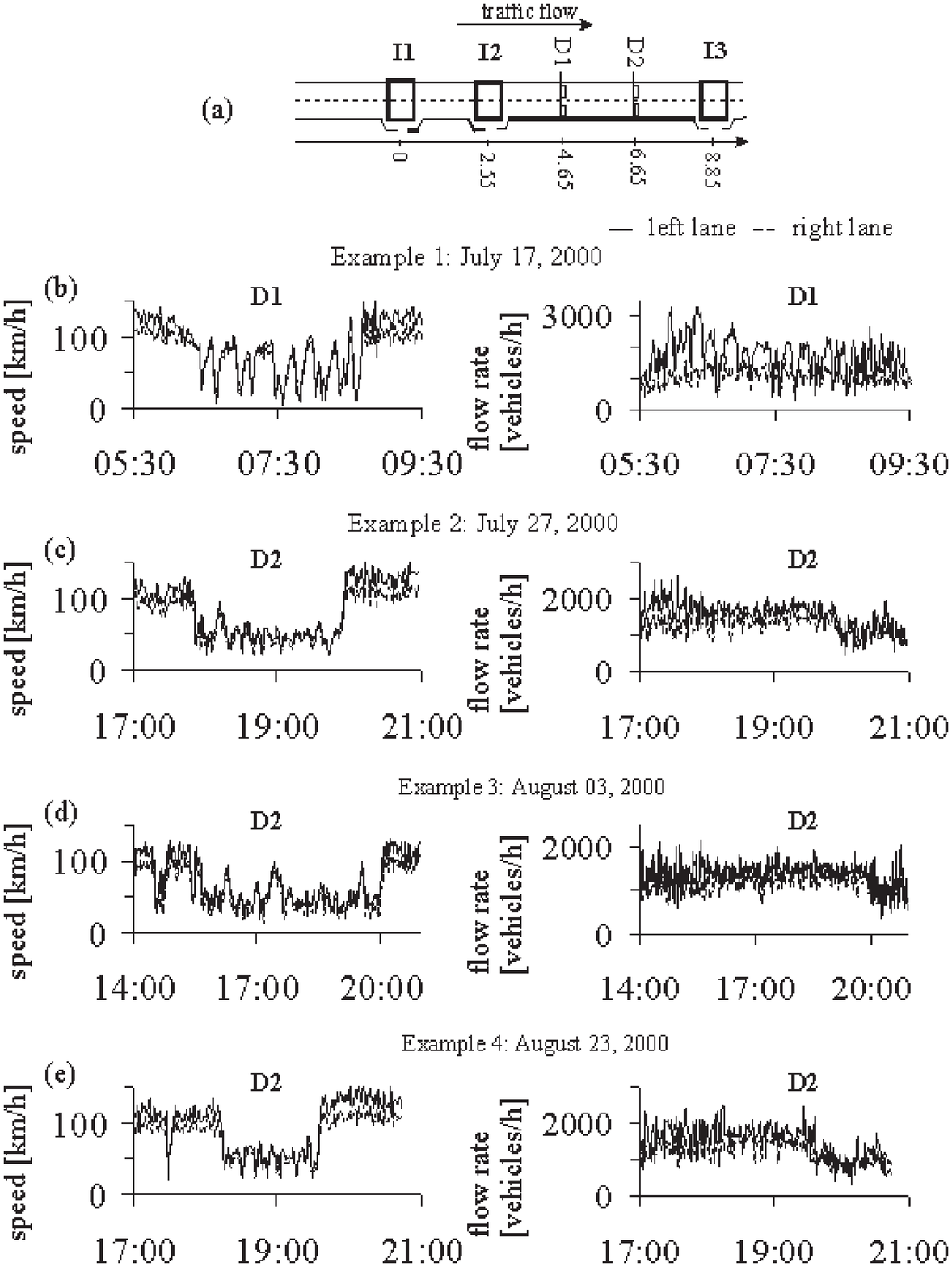}
\caption{Macroscopic characteristics of empirical single vehicle data: (a) -- Sketch of detector arrangement.
(b) -- Local traffic dynamics (one-minute average data) of the speed (left) and flow rate (right) on four different days
in both freeway lanes.
 \label{Single_Fig} } 
\end{center}
\end{figure*}

Single vehicle data have been measured from June 01, 2000 till September, 30, 2000 on a two-lane section of the freeway A92-West
with two sets of double induction loop detectors (D1 and D2) between intersections $\lq\lq$AS Freising S\"{u}d" (I2)
and  $\lq\lq$AK Neufahm" (I3)
near  Münich Airport in Germany (intersection I1: $\lq\lq$AS Flughafen" in 
Fig. ~\ref{Single_Fig} (a)). 
 Each    detector set consists of two detectors for the left (passing) and right lanes.
A detector registers a vehicle by producing  a   current electric pulse whose duration $\Delta t_{i}$
is related to the time taken by the vehicle to traverse the induction loop. This enables us to calculate 
the gross time gap between two vehicles $i$ and $i+1$ that have passed the loop one after the other
 $\tau^{\rm (gross)}_{i,i+1}$. There are two detector loops in each detector.
The distance between these loops is constant. This enables us to calculate the individual vehicle speed $v_{i}$,
the vehicle length $d_{i}=v_{i}\Delta t_{i}$, as well as the net time gap (time headway) between the vehicles $i$ 
 and $i+1$: $\tau_{i,i+1}=\tau^{\rm (gross)}_{i,i+1}-\Delta t_{i}$.
 
Local dynamics of the average speed and flow rate (one-minute average data) for four typical
days at which congested traffic have been observed are shown in Fig. ~\ref{Single_Fig} (b). The local
congested traffic  dynamics from July 17, 2000 is designated below as $\lq\lq$example 1", whereas
the dynamics from July 27, August 03, and 27, 2000 are designated as $\lq\lq$examples 2--4", respectively.

From   consideration of the flow rate for the examples 1--4, it can be seen
that even if within the local  dynamics for the examples 2--4  there are moving jams, then these jams should mostly be narrow moving jams,
i.e., these examples should correspond to the synchronized flow phase. 
Indeed, there are no  drops in the flow rate, which are typical for wide moving jams~\cite{KernerBook},
 during the whole time of congested traffic existence in the examples 2--4.
In contrast, in the example 1  there are short-time drops both in the average speed and flow rate  
  within the local congested dynamics (example 1, left in Fig. ~\ref{Single_Fig} (b)). Such  drops are usually typical for 
  wide moving jams~\cite{MovingJams}. However, from these {\it macroscopic} data (example 1, Fig. ~\ref{Single_Fig} (b))
  measured at a single location only
  we cannot make a conclusion with certainty
  whether these drops are associated with the wide moving jam phase or not. Nevertheless,
  this conclusion can  be made, if a microscopic
    criterion for the traffic phases presented below is applied.

\subsection{Traffic Flow Interruption Effect as Microscopic Criterion for Wide Moving Jam}
\label{Interruption}

The characteristic wide moving jam feature to propagate through a bottleneck while maintaining the downstream jam velocity, which
distinguishes the wide moving jam from synchronized flow (in accordance with the macroscopic spatiotemporal
objective criteria for the traffic phases in congested traffic),
can be explained by  a traffic flow discontinuity within a wide moving jam.
Traffic flow is interrupted by the wide moving jam: There is no influence of the inflow into the jam on the jam outflow.
For this reason, the mean downstream   jam front velocity and the jam outflow exhibit fundamental characteristic features, which
  do not depend on the jam inflow~\cite{KernerBook}. 
  A difference between the jam inflow and the jam outflow changes the jam width only.
  This  {\it traffic flow interruption   effect} is a general effect for each wide moving jam. 

The jam outflow becomes independent of the jam inflow, when the traffic flow interruption effect within a moving jam
 occurs. This is realized, when due to a very low  vehicle  speed or a vehicle standstill
the maximum gross time headway $\tau^{\rm (gross)}_{\rm max}$ between  two vehicles within the   jam is 
 considerably longer than the mean time
delay  $\tau^{\rm (ac)}_{\rm del}$ in vehicle acceleration  at the downstream jam front  
from a standstill state~\cite{LongVehicles}:
\begin{equation}
\tau^{\rm (gross)}_{\rm max}\gg \tau^{\rm (ac)}_{\rm del}.
\label{GrossDel}
\end{equation}   
 Indeed, the time delay $\tau^{\rm (ac)}_{\rm del}$
determines  the jam outflow~\cite{TimeDelay}.
Under the condition (\ref{GrossDel}), there are at least several vehicles within the jam that    are
in a standstill or if they are still moving, it is only with a negligible low speed in comparison with the speed
in the jam inflow and outflow. These vehicles separate vehicles accelerating at the downstream jam front from
vehicles  decelerating at the upstream jam front: The inflow into the jam has no influence on the jam outflow. Then
the jam outflow is fully determined by vehicles accelerating  at the downstream jam front.
As a result, the mean downstream jam front velocity is equal to $v_{\rm g}=-1/(\tau^{\rm (ac)}_{\rm del}\rho_{\rm max})$~\cite{KernerBook} 
($\rho_{\rm max}$ is the density within the jam)
regardless of whether there are  bottlenecks or other complex traffic states 
on the freeway. In other words, this jam propagates through a bottleneck
while maintaining the mean downstream jam front velocity $v_{\rm g}$, i.e., in accordance with
the macroscopic objective spatiotemporal criteria for the traffic phases in congested traffic
(Sect.~\ref{Intr}) this jam is a wide moving jam.

  Thus, the traffic flow interruption   effect
can be used as a criterion to distinguish between the synchronized flow and wide moving jam phases in single vehicle data.
This is possible 
even  if  data is measured at a single freeway location. This enables us to find
dependence of single vehicle characteristics on different local microscopic congested pattern features (Sect.~\ref{Single_Dep_Sect}).

\begin{figure*}
\begin{center}
\includegraphics[width=12 cm]{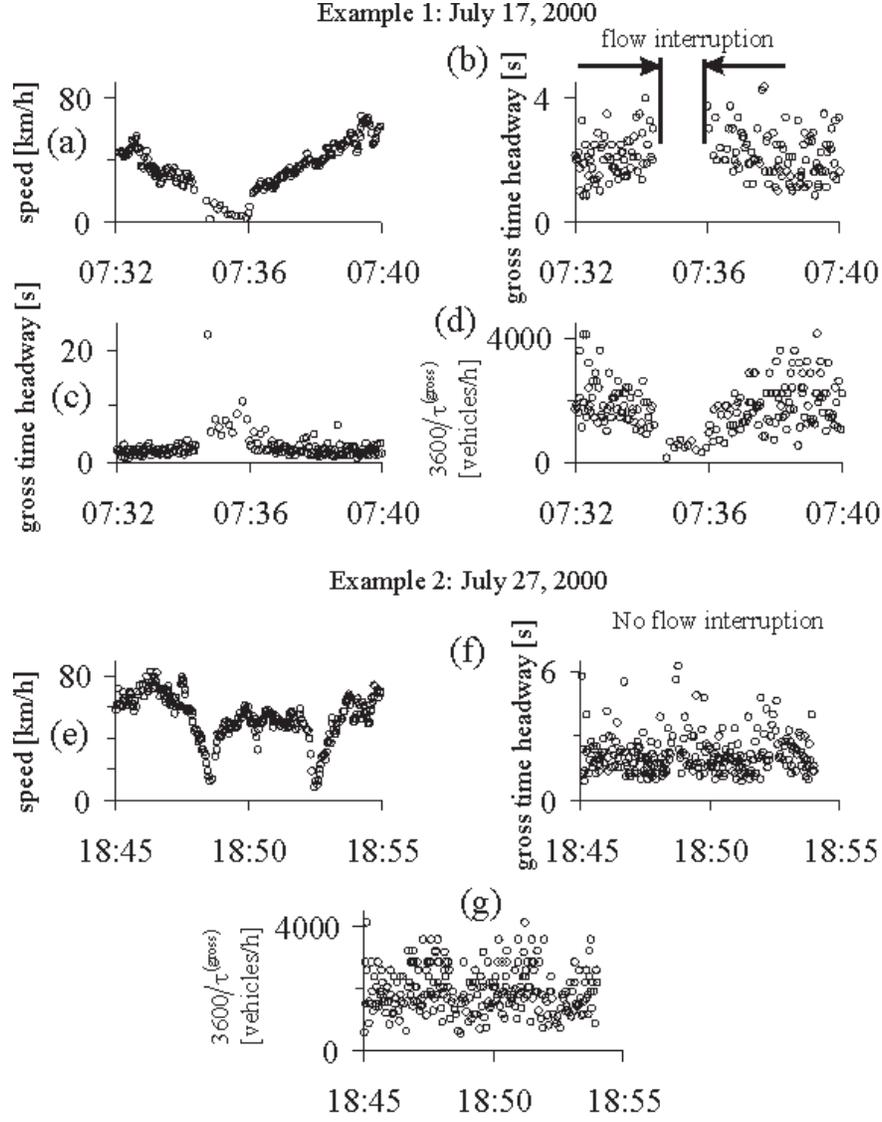}
\caption{Microscopic criterion for wide moving jam:
(a--d) -- Empirical single vehicle data for speed within a wide moving jam in the left lane (a) and
the associated time distributions of gross time headways $\tau^{\rm (gross)}$ for different scales of the time headways (b, c) 
and of the value $3600/\tau^{\rm (gross)}$ (d) related to the example 1 in Fig.~\ref{Single_Fig} (b).
(e--g) -- Empirical single vehicle data for speed within a sequence of two narrow moving jams in the left lane (e) and
  the associated time distributions of $\tau^{\rm (gross)}$ (f)
  and of $3600/\tau^{\rm (gross)}$ (g)
  related to the example 2 in Fig.~\ref{Single_Fig} (b).
 \label{Iterr_Fig} } 
\end{center}
\end{figure*}

The interruption of traffic flow within a moving jam shown in 
Fig.~\ref{Iterr_Fig} (a) is clearly seen in the time-dependences of gross time headways
$\tau^{\rm (gross)}$ (Fig.~\ref{Iterr_Fig} (b, c))
and of the value $3600/\tau^{\rm (gross)}$ (Fig.~\ref{Iterr_Fig} (d)).
Before and after the jam has passed the detector D1 (due to the jam upstream propagation~\cite{MovingJams})
there are many vehicles that traverse the induction loop of the detector. Within the jam, there are no vehicles traversing the detector
 with   synchronized flow
 gross time headways $< 4$ sec during a time interval (this time interval is labeled $\lq\lq$flow interruption" in Fig.~\ref{Iterr_Fig} (b)),
 when the speed within the jam is approximately   zero (Fig.~\ref{Iterr_Fig} (a)).
This means that traffic flow is discontinuous within the moving jam, i.e., this moving jam is associated with the
 wide moving jam   phase. If all gross time headways for the time interval are considered (Fig.~\ref{Iterr_Fig} (c)),
 then it can be seen that the $\lq\lq$real" duration of  flow interruption within the wide moving jam, i.e., when
 vehicles do not move through the detector at all,
  is equal to approximately 20 sec: There is a vehicle with
  the gross  time headway $\approx$ 20 sec. Later, some vehicles, which are within the jam, exhibit gross time headways 
  about 5 sec or longer.
 The latter  can be explained by moving blanks within the jam (see Sect.11.2.4 in~\cite{KernerBook}). 
 This is correlated with the result that 
 only after the jam (at $t>$ 7:36) has passed the detector,  
  synchronized flow gross time headways (1--4 sec) are observed (Fig.~\ref{Iterr_Fig} (c)).
 Similar results are found for other moving jams in single vehicle data.

 In examples 2--4 (Fig. ~\ref{Single_Fig} (b)), there also are many moving jams during the time intervals of congested traffic.
 For example, the speed  within  two moving jams in Fig.~\ref{Iterr_Fig} (e) (the example 2)
 is also very low.
 Nevertheless,
in contrast with the wide moving jam in Fig.~\ref{Iterr_Fig} (a) rather than wide moving jams  these moving jams should be classified as
narrow moving jams~\cite{KernerBook}. This is because there are no traffic flow interruptions within  
these moving jams (Fig.~\ref{Iterr_Fig} (f, g)). Indeed,
upstream, downstream of the jams, and within the jams there are many vehicles that traverse the induction loop of the detector:
There is no qualitative difference in the time-dependences of gross time headways for different
time intervals associated with these narrow jams and in traffic flow upstream or downstream of the  jams (Fig.~\ref{Iterr_Fig} (f)).

This can be explained if it is assumed that each vehicle, which meets  a narrow moving  jam,  must
decelerate down to a very low speed within the jam, 
 can nevertheless accelerate later almost without any time delay within the jam: The narrow moving jam consists of 
 upstream and downstream jam fronts only. Within the upstream front vehicles   must decelerate. However,
 they then can  accelerate almost immediately at the downstream jam front.  These assumptions are confirmed by single vehicle data shown in Fig.~\ref{Iterr_Fig} (f, g), in which
 time intervals between different measurements of gross time headways and for the value $3600/\tau^{\rm (gross)}$ 
  for different vehicles exhibit the same behavior
 away and within the jams.
Thus, regardless of these narrow moving jams traffic flow is not discontinuous, i.e., the narrow moving jams belong indeed to the 
synchronized flow traffic phase.

This single vehicle analysis enables us to conclude
 that congested traffic in the example 1 is associated with a sequence of   wide moving jams
 propagating in synchronized flow. In contrast, 
congested traffic in the examples 2--4 is mostly associated with different states of the synchronized flow  phase.

\subsection{Dependence of Single Vehicle Characteristics on  Local Congested Pattern Features}
\label{Single_Dep_Sect}

\subsubsection{Time Headway Distributions}

\begin{figure*}
\begin{center}
\includegraphics[width=12 cm]{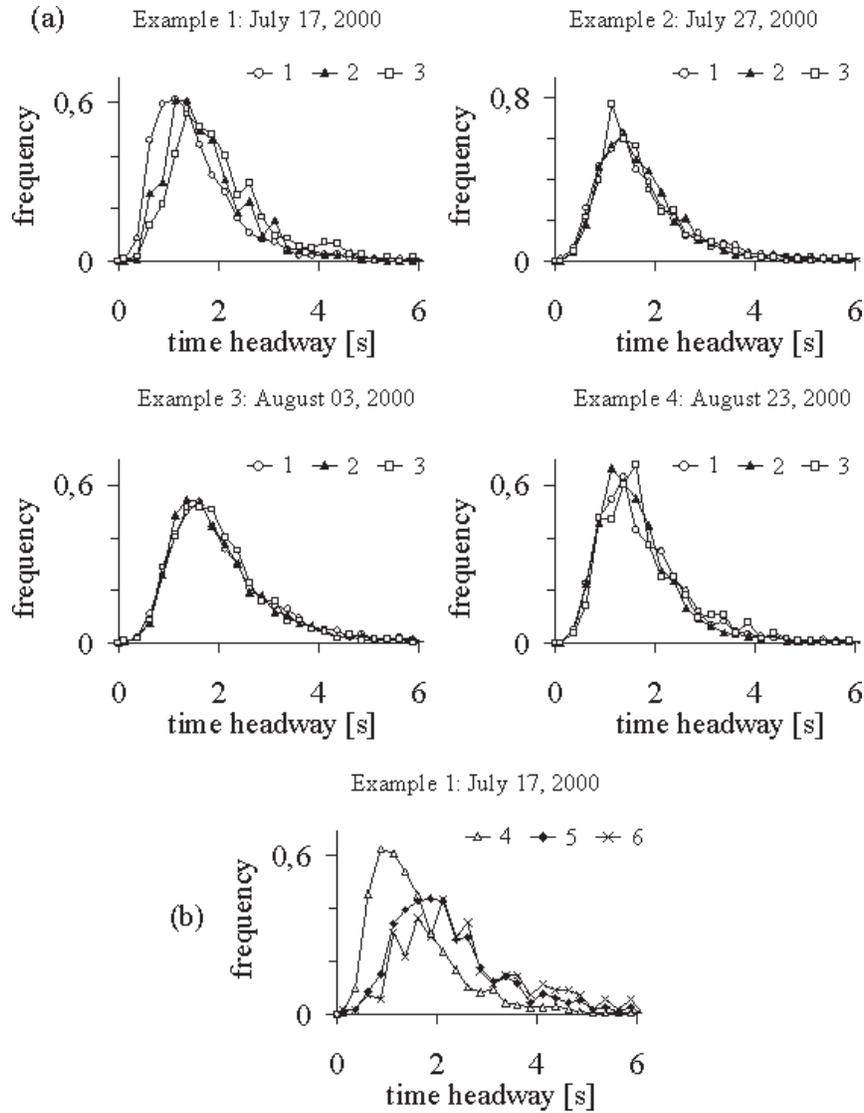}
\caption{Empirical time headway distributions: (a) -- For the examples 1--4 in Fig.~\ref{Single_Fig} (b)
for different density ranges: curve 1 -- 20 $<\rho<$ 30, curve 2 -- 30 $<\rho<$ 40, curve 3 --
40 $<\rho<$ 50 vehicles/km.
(b) -- Time headway distributions associated with the example 1 for different
speed ranges: curve 4 -- $v>$ 50, curve 5 -- $v<$ 30, curve 6 --
$v<$ 20 km/h. Left lane.
 \label{TimeHeadways} } 
\end{center}
\end{figure*}

Although local traffic dynamics in the example 1 is qualitatively different from the examples 2--4, the related time 
headway distributions found for different density ranges  are qualitatively the same (Fig.~\ref{TimeHeadways} (a)). 
Moreover, the distributions are both qualitatively and even approximately 
quantitatively  the same as those found  for congested traffic on different freeways in various 
countries (e.g.~\cite{Bovy,Neubert,Tilch2000,Knospe2002}). Thus, based only  on these time
 headway distributions it is not possible  to distinguish different synchronized flow local dynamics
in the examples 2--4 one from another and also from the example 1 with wide moving jams.

However, the time headway distributions for the example 1 with wide moving jams
exhibit some peculiarities: If 
  time headway distributions for vehicles whose speed $v>$ 50 km/h, $v<$ 30 km/h,
and $v<$ 20 km/h are drawn separately, then it turns out that the lower the speed, the more shifted become
 the time headway distributions to longer time headways (Fig.~\ref{TimeHeadways} (b)).
As should be expected, the time headway distributions for the vehicles away from the jams (with $v>$ 50 km/h)
 are almost the same as those for   synchronized flow without a wide moving jam sequence
(examples 2--4). In contrast,  
vehicles within wide moving jams 
exhibit appreciable longer time headways, which are the longer, the lower the speed. 
As a data analysis shows, these time headways are mostly associated with moving blanks within the jams.
However, even these differences in the time headway distributions 
 say nothing about jam duration,  speed distributions between the jams,
 and many other 
spatiotemporal congested traffic characteristics. 

We can conclude that the microscopic characteristic of congested traffic 
$\lq\lq$time headway distribution" cannot be used for  clear distinguishing    spatiotemporal
congested pattern features.  
This is because within time headway distributions 
most of spatiotemporal traffic characteristics are
averaged. Because the sum of wide moving jam duration in the example 1 is  shorter than the synchronized flow duration,
it is almost impossible to distinguish much longer time headways associated with moving blanks within the jams.
Only when the share of the jams in the time headway distribution  increases
(Fig.~\ref{TimeHeadways} (b), $v<$ 20 km/h), the effect of  moving blanks can 
be identified.

\subsubsection{Optimal Velocity (OV) Functions}

\begin{figure*}
\begin{center}
\includegraphics[width=12 cm]{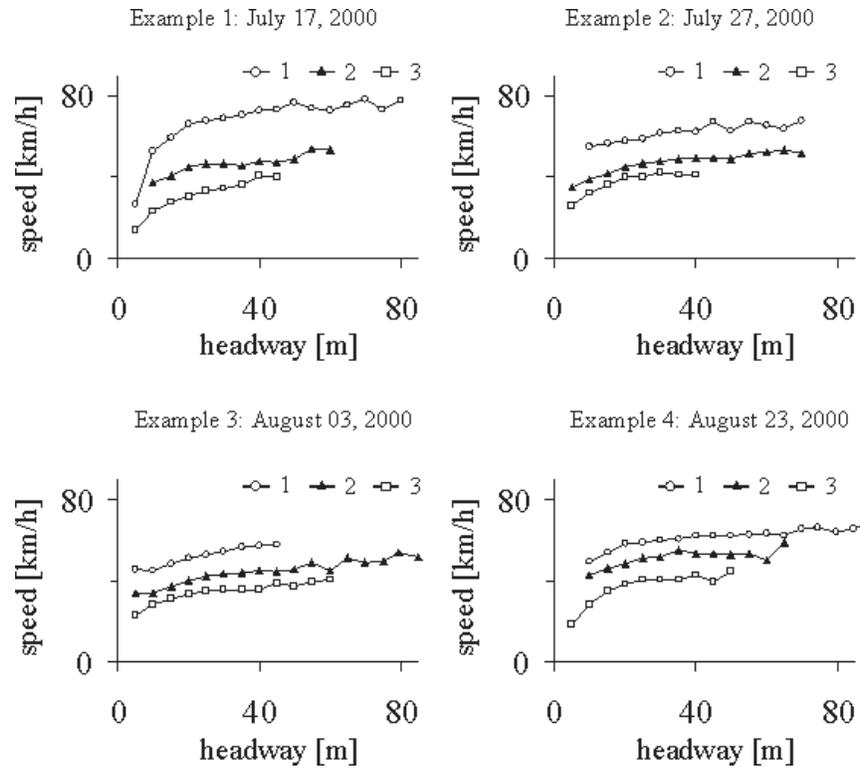}
\caption{Empirical OV functions for the examples 1--4   in Fig.~\ref{Single_Fig} (b)
for different density ranges:
Curves 1 -- $20<\rho<30$, curves 2 -- $30<\rho<40$, curves 3 -- $40<\rho<50$ vehicles/km. Left lane.
 \label{OV_Func} } 
\end{center}
\end{figure*}

OV functions are space gap (headway) dependencies of the mean vehicle speed
calculated for different given density ranges~\cite{Neubert}.
The OV functions found  do not exhibit some qualitative differences for the different examples 1--4
of local congested traffic dynamics (Fig.~\ref{OV_Func}). They are qualitatively the same as the OV functions
first derived for aggregated single vehicle data in~\cite{Neubert}: At smaller headways the speed increases with headways
considerably, whereas for larger headways there is a saturation effect for the speed growth.

We can conclude that   microscopic   
 OV functions cannot also be used for  clear distinguishing    spatiotemporal
congested pattern features. This is because within OV functions 
 many of the spatiotemporal traffic characteristics are
averaged.

\subsection{Spatiotemporal Macroscopic Congested Patterns}
\label{STEmpTest_Sect}

In Sects.~\ref{STEmpTest_Sect} and~\ref{Spatial_Emp_M}, a brief discussion of 
empirical macroscopic spatiotemporal traffic features considered in~\cite{Kerner2002B,KernerBook}
is made  that is necessary for the empirical test of a microscopic three-phase traffic theory. 

Results of empirical investigations reviewed in the book~\cite{KernerBook} enable us to conclude that
there are a huge number of different  congested traffic patterns,
whose spatiotemporal structure depends on type, feature, and location of a freeway bottleneck(s),  on
other peculiarities of a freeway network, as well as on traffic demand, weather, and other traffic conditions.
 At  first glance all these look like very different patterns, however, it turns out  they  exhibit
clear common features and characteristics~\cite{KernerBook}.
Empirical investigations  of data  measured over many days and years on freeways in various countries
show that common traffic phenomena and characteristics of congested patterns, which are most frequent observed, are as follows. (i) GP emergence and evolution that occur at an on-ramp bottleneck (Sect.~\ref{GP_Emp_Sect}).
(ii) Complex congested pattern emergence and transformation
 that occur at two adjacent off- and on-ramp bottlenecks (Sect.~\ref{WSP_Emp_Sect}).

\subsubsection{General Pattern at On-Ramp Bottleneck}
\label{GP_Emp_Sect}

In Fig.~\ref{150496_Pattern}, general pattern (GP) formation   at an on-ramp bottleneck  is shown~\cite{KernerBook}. Firstly, 
an F$\rightarrow$S transition occurs at the bottleneck (up-arrow  at detector D6). Synchronized flow
propagates upstream (up-arrows $S$ at detectors D5--D4), whereas free flow remains downstream (detector D7):
The downstream front of synchronized flow is fixed at the bottleneck. 

Upstream of the bottleneck the pinch effect in synchronized flow is realized:
The speed decreases (detector D5) and density increases. Narrow moving jams emerge spontaneously
in the pinch region in synchronized flow (detectors D5--D4). These jams propagate upstream. Some of the jams transform into wide moving jams:
The region of wide moving jams propagating upstream is formed
(detectors D3--D1). 

\subsubsection{Complex Pattern Emergence and Transformation on Freeways with two Different Adjacent Bottlenecks}
\label{WSP_Emp_Sect}

Characteristic features of complex pattern emergence and transformation are often observed at
a freeway section with two adjacent bottlenecks: An off-ramp bottleneck that is downstream 
and an on-ramp bottleneck that is upstream~\cite{KernerBook}.
In the example in Fig.~\ref{230301_GP},
firstly a widening SP (WSP) at the off-ramp bottleneck emerges.  Moreover, an expanded pattern (EP) appears, in which synchronized flow affects
both bottlenecks. The EP appears after   synchronized flow of the initial WSP covers the on-ramp bottleneck.
Secondly, the
 WSP transforms into an GP at the off-ramp bottleneck.
 
After the synchronized flow covers the on-ramp bottleneck, wide moving jams begin to emerge
downstream of the on-ramp bottleneck  within the initial WSP: Rather than the WSP remaining, an GP appears at the off-ramp bottleneck, i.e.,
congestion emergence at the upstream bottleneck leads to intensification of congestion at the downstream bottleneck.
Wide moving jams (labeled 1--4 in Fig.~\ref{230301_GP} (b))
that emerge within the GP, i.e., between the off-ramp and on-ramp bottlenecks, propagate through the on-ramp
bottleneck while maintaining the mean velocity of the jam downstream front.

\subsection{Spatial Dependence of Empirical Macroscopic Traffic Flow Characteristics}
\label{Spatial_Emp_M}

Because a qualitative behavior of the congested traffic dynamics  
strongly depends on a chosen detector location within a pattern, local pattern characteristics are functions
on the spatial co-ordinate. 
In particular,
the fundamental diagram depends both on the pattern type and
on a location within a congested pattern at which the flow rate and speed are measured. 
The fundamental diagram consists of the branches for free flow (curves $F$)
and congested traffic (curves $C$ in Fig.~\ref{150496_FD}).

In the case of the GP  at the on-ramp bottleneck (Fig.~\ref{150496_Pattern} (b, c)),
at the locations D6 and D5 the braches $C$ for congested traffic are associated with  synchronized flow only.
In synchronized flow at D6, 
the branch $C$ has a positive slope in the flow--density plane. This behavior is changed for the pinch region of synchronized flow
 (D5). Due to narrow moving jam emergence in the pinch region
at  greater densities the branch $C$ has a slightly negative slope, whereas at smaller densities of synchronized flow
it has a slightly positive slope, i.e., there is a maximum on the branch $C$. However, 
this maximum is very weak: The flow rate   does not approximately depend on density within the pinch region of the GP.

This   is explained
 in Fig.~\ref{150496_FD} (b), in which empirical data from the   time interval  
 06:45---8:00 of the 
 strong congestion condition are used only. In this case, in the pinch region (at D5) the part on the curve $C$ with the positive slope 
 disappears and the flow rate is  almost
 independent on density (only at greater density there is a slight decrease in flow rate  due to narrow moving jam emergence
 in the pinch region).
 This also means that the part of the curve $C$ with the positive slope at D5 is associated with
the time interval after 8:00, when  strong congestion changes to weak congestion.

At the locations D4 and D3, wide moving jams begin to form. For this reason,
 the branch $C$ asymptotically approaches the line $J$ with greater density~\cite{KernerBook}. 
In the region of wide moving jams (D2, D1), 
the branch $C$ lies on the line $J$
with greater densities associated with the  outflow from wide moving jams. 
The positive slope of the branch $C$ at D4--D1 (as at the location D5) at smaller densities 
is associated with the time interval
after 8:00
when the flow rate upstream of the on-ramp bottleneck $q_{\rm in}$ and the flow rate to the on-ramp 
$q_{\rm on}$ decrease appreciably and 
synchronized flow is formed in which the flow rate is an increasing density function.
If data related to this time interval is not taken into account, then for the region of wide moving jams the whole branch 
for congested traffic $C$ lies on the line $J$ only~\cite{Wide_J}.

Because the empirical fundamental diagram for congested traffic depends considerably 
on the spatial location within
the pattern,   
some $\lq\lq$global", aggregated, and other averaged  fundamental diagrams, which are often used for
 an empirical test of traffic flow models (e.g.,~\cite{Knospe2004}), cannot answer the question whether
 a model can show and predict  spatiotemporal traffic phenomena.

Another important macroscopic empirical characteristic whose spatial dependence
should be shown by a traffic flow model is the speed correlation function (see Fig. 12.8 in~\cite{KernerBook}
associated with Fig.~\ref{150496_Pattern} (a)).
The period of the speed correlation function $T_{\rm c}$ has a minimum value (about 5 min for the GP)
within the pinch region.  
Propagating upstream some of these jams disappear and other transform into wide moving jams.
As a result, $T_{\rm c}$ increases in the upstream direction reaching the maximum value
(about 10 min for the GP), when the region of wide moving jams
is formed completely. 

\section{Microscopic and macroscopic traffic model characteristics}
\label{MST_Sect}

In numerical simulations of a microscopic three-phase traffic theory presented below,
single vehicle model data, which should be compared with empirical microscopic traffic characteristics
of Sects.~\ref{Interruption} and~\ref{Single_Dep_Sect},
are related to  spatiotemporal congested model patterns.
In turn, these patterns as well as their features should correspond to empirical observations.
For this reason, before we compare {\it microscopic} model and empirical results,
it is necessary to prove whether {\it macroscopic}
features of these spatiotemporal congested model patterns are 
associated with empirical  congested patterns of Sect.~\ref{STEmpTest_Sect}.

To reach this goal, firstly empirical
time-dependence of traffic demand and drivers' destinations (whether a vehicle 
leaves the main road to an off-ramp or it further follows the main road) associated with
these empirical macroscopic patterns are used in model simulations at the upstream model boundaries
of the main road and of an on-ramp. At   downstream model boundary conditions for vehicle freely leaving  
a modeling freeway section(s)
are given, which are not associated with congested traffic propagating upstream.
Then spatiotemporal congested model patterns emerge, develop, and dissolve in this open freeway model 
 with the same types of bottlenecks as those in empirical observations (Sect.~\ref{ST_Sect}).
Finally, single vehicle model data are analysed for different locations within these patterns
(Sects.~\ref{Interrup_S} and~\ref{Model_Single_Cr}).
 
 Note that a study of statistical   features of free flow satisfactory  investigated  both  empirically and theoretically
in many previous works (see references in e.g.,~\cite{Sch,Helbing2001}) is beyond the scope of the article. Because the aim of the paper
is to study congested pattern features, we  can use  the models
 of Ref.~\cite{KKl2003A,KKW,KKl2004AA}
 in which a very simplified model of free flow is used. 
 
 We use   models of bottlenecks and model parameters
 of Ref.~\cite{KKl2003A,KKW,KKl2004AA} (see a detailed consideration of the models, their physics and parameters in
  Sects. 16.2, 16.3, and 20.2 of the book~\cite{KernerBook}). When some other
 model parameters are   used, they will be given in   figure captions.
 
\subsection{Macroscopic Spatiotemporal Features of Congested Pattern Evolution}
\label{ST_Sect}

\subsubsection{Occurrence and Evolution of General Pattern under Strong Congestion}
 \label{GP_SC}
 
\begin{figure*}
\begin{center}
\includegraphics[width=12 cm]{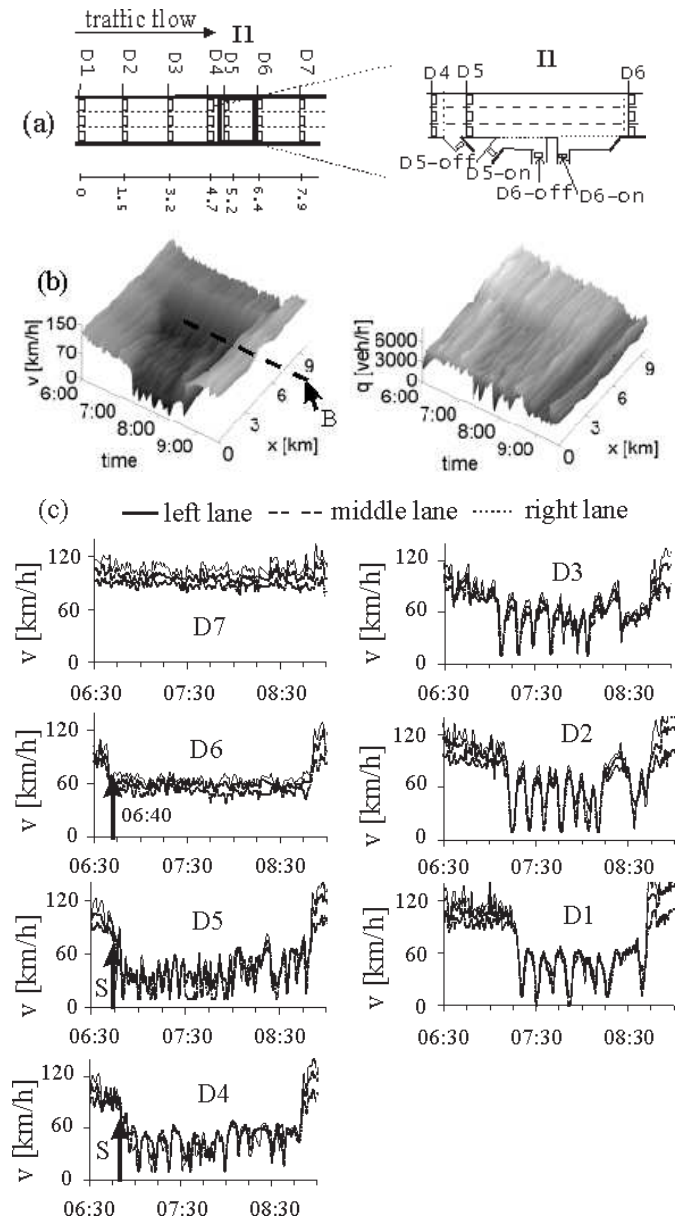}
\caption{GP upstream of  the bottleneck at the effectual on-ramp D6
on a section
of the freeway A5-South: (a) -- Sketch of the freeway section (left) with 
detector arrangement within the intersection $I1$ (right).
(b) -- Vehicle speed averaged  across  all lanes (\emph{left}) 
and   total flow rate across the freeway (\emph{right}) in space and time.
(c) -- Vehicle speed in different freeway lanes within the GP at different detectors. Data from April 15, 1996. 
 Explanation of the section and the on-ramp bottleneck at D6 see in~\cite{KernerBook}.
  \label{150496_Pattern} } 
\end{center}
\end{figure*}

 For numerical simulation of empirical GP occurrence and evolution at an on-ramp bottleneck (Fig.~\ref{150496_Pattern} (b, c)),
 a model of two-lane freeway with an on-ramp bottleneck  is used (Sect. 16.2 in~\cite{KernerBook}).
 At the upstream boundary of the main road, the time-dependence of the flow rate $q_{\rm in}(t)$ associated with empirical
 data measured at the detector D1  has been applied (Fig.~\ref{GP} (a, b)). Empirical detector measurements (Fig.~\ref{150496_Pattern} (b, c)) are compared with simulated
 results at the same virtual detector locations D1--D6 
 as those in empirical data.   As wide moving jams are observed during the time interval 
 7:15--9:00  at the farthest available upstream detector D1,
 the flow rate $q_{\rm in}(t)$ is approximated  by a line.
 The flow rate to the on-ramp $q_{\rm on}(t)$ used in the simulations is  taken from measurements in the on-ramp lane.

In simulations, GP emergence and evolution (Fig.~\ref{GP} (c, d)) 
are qualitatively and quantitatively the same as those in empirical
 observations (Fig.~\ref{150496_Pattern}). In particular,  
in accordance with the empirical study, the following main effects are found:
 
(i) {\it A local first-order F$\rightarrow$S transition at the bottleneck} (labeled $S$ at D6,  Fig.~\ref{GP} (e)).
Over time traffic demand ($q_{\rm on}(t)$ and $q_{\rm in}(t)$) in an initial free flow at the on-ramp bottleneck increases. 
Whereas at D7 just downstream of the bottleneck and just upstream of the bottleneck at D5
free flow remains, at D6 within the merging region of the on-ramp
an abrupt decrease in speed is observed.

(ii) {\it A probabilistic nature} of the F$\rightarrow$S transition: In different realizations performed 
at the same $q_{\rm on}(t)$ and $q_{\rm in}(t)$,
 an F$\rightarrow$S transition can occur at different flow rates at the bottleneck.

(iii) {\it Upstream propagation of a wave of induced F$\rightarrow$S transitions.} Whereas free flow further remains at D7 downstream of the
bottleneck, F$\rightarrow$S transitions are induced by
 upstream propagation of the upstream front of synchronized flow 
   (labeled $S$ in  Fig.~\ref{GP} (e), D5 and D4).

(iv) {\it Pinch region formation with narrow moving jam emergence within synchronized flow.}
A self-compression of synchronized flow upstream of the bottleneck (D5, D4) is observed:
Average speed decreases and  density increases. In this pinch region, narrow moving jams
emerge and grow in amplitude propagating upstream.

(v) {\it Formation of wide moving jams.} Some of narrow moving jams transform
into wide moving jams (D3). As a result, a region of wide moving jams upstream of the pinch region is formed
(D3--D1). Wide moving jams propagate upstream while maintaining the mean velocity of their
downstream front. In the      wide moving jam outflow either free flow or synchronized flow can occur.
In the first case, in the flow--density plane the line $J$
reaches free flow region (Fig.~\ref{GP} (f)). In the second case, 
the left co-ordinates of the line $J$ are related to the average speed and flow rate 
in  synchronized flow (Fig.~\ref{GP} (g)).

(vi) {\it Strong congestion conditions.}
During the time before $t\approx$ 8:00, strong congestion conditions are
realized in the pinch region. The average flow rate (10 min averaging time interval) displays only
minor changes in the vicinity of the limit (minimum) flow rate $q^{\rm (pinch)}_{\rm lim}$, which does not depend on traffic demand,
specifically when the flow rates $q_{\rm on}(t)$ and $q_{\rm in}(t)$) increase;
the frequency of narrow moving jam emergence reaches the maximum possible value for chosen model parameters.

(vii) {\it GP evolution.}  When the flow rates $q_{\rm on}(t)$ and $q_{\rm in}(t)$
begin to decrease, the initial GP under the strong congestion condition transforms into an GP under the weak congestion condition.
In this case, the average speed in the pinch region 
increases and the average flow rate  ceases to be a self-sustaining value
that is close to
 $q^{\rm (pinch)}_{\rm lim}$;
the frequency of narrow moving jam emergence in the pinch region decreases.

(viii) {\it Return S$\rightarrow$F transitions.} When the flow rates $q_{\rm on}(t)$ and $q_{\rm in}(t)$ further decrease,
 return S$\rightarrow$F transitions firstly occur upstream of the bottleneck and later at the bottleneck.
As a result, the congested pattern dissolves and free flow returns at the bottleneck.

(ix)  {\it  Hysteresis Effects.}  
At each detector location F$\rightarrow$S   and return S$\rightarrow$F transitions cause hysteresis effects
in the flow--density plane~\cite{Hysteresis}.

 \subsubsection{WSP Emergence and its Transformation into GP and EP}
 \label{WSP_GP_EP}
 
\begin{figure*}[t]
\begin{center}
\includegraphics[width=12 cm]{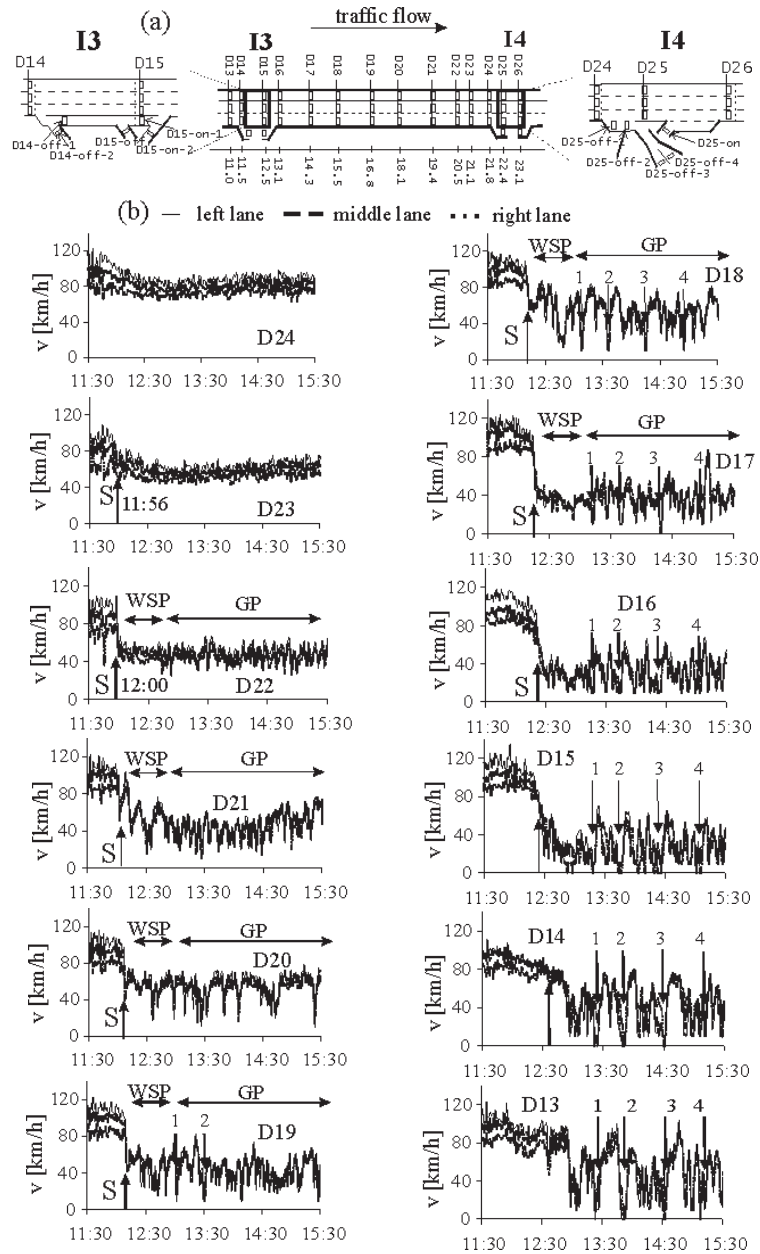}
\end{center}
\caption[]{Congested pattern evolution on a section on the freeway A5-North with off-ramp (at D25-off) and on-ramp 
(at D16) bottlenecks: (a) -- Sketch of the section (middle) with detector arrangement within the intersection $I3$ (left)
and $I4$ (right). (b) -- Vehicle speed over time shown at different detectors. 
 Explanation of the section and the on-ramp and off-ramp bottlenecks   see in~\cite{KernerBook}.
}
\label{230301_GP}
\end{figure*}

\begin{figure*}
\begin{center}
\includegraphics[width=12 cm]{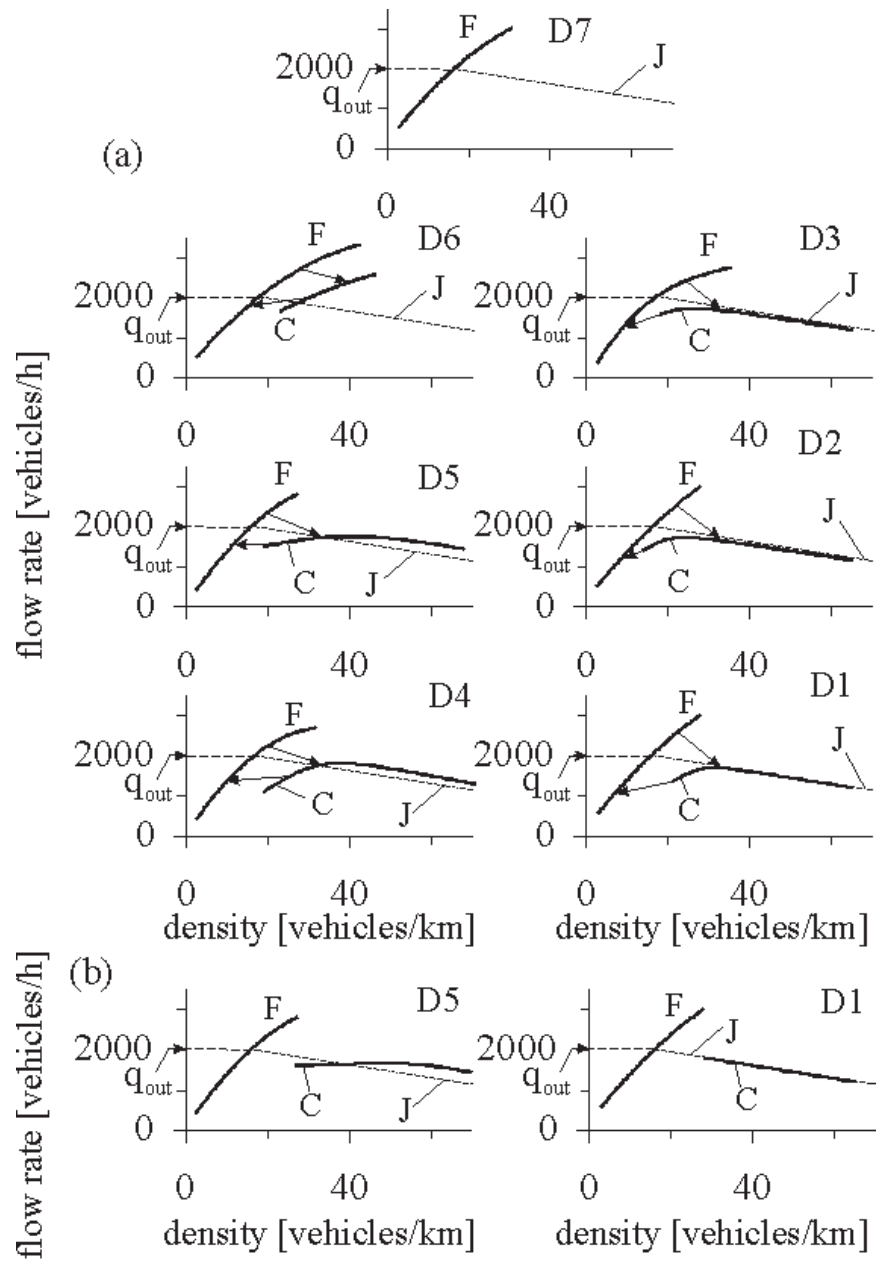}
\caption{Empirical fundamental diagram at different freeway locations for the left freeway
lane related to the GP 
in Fig.~\ref{150496_Pattern}: (a) -- All data  is used~\cite{KernerBook}.
(b) -- Only data before 8:00 is used. Dotted line shows the line $J$, which represents
the upstream propagation of the wide moving jam downstream front; $q_{\rm out}$ is the flow rate in free flow formed by
the wide moving jam outflow.}
\label{150496_FD}
\end{center}
\end{figure*}

\begin{figure*}
\begin{center}
\includegraphics[width=12 cm]{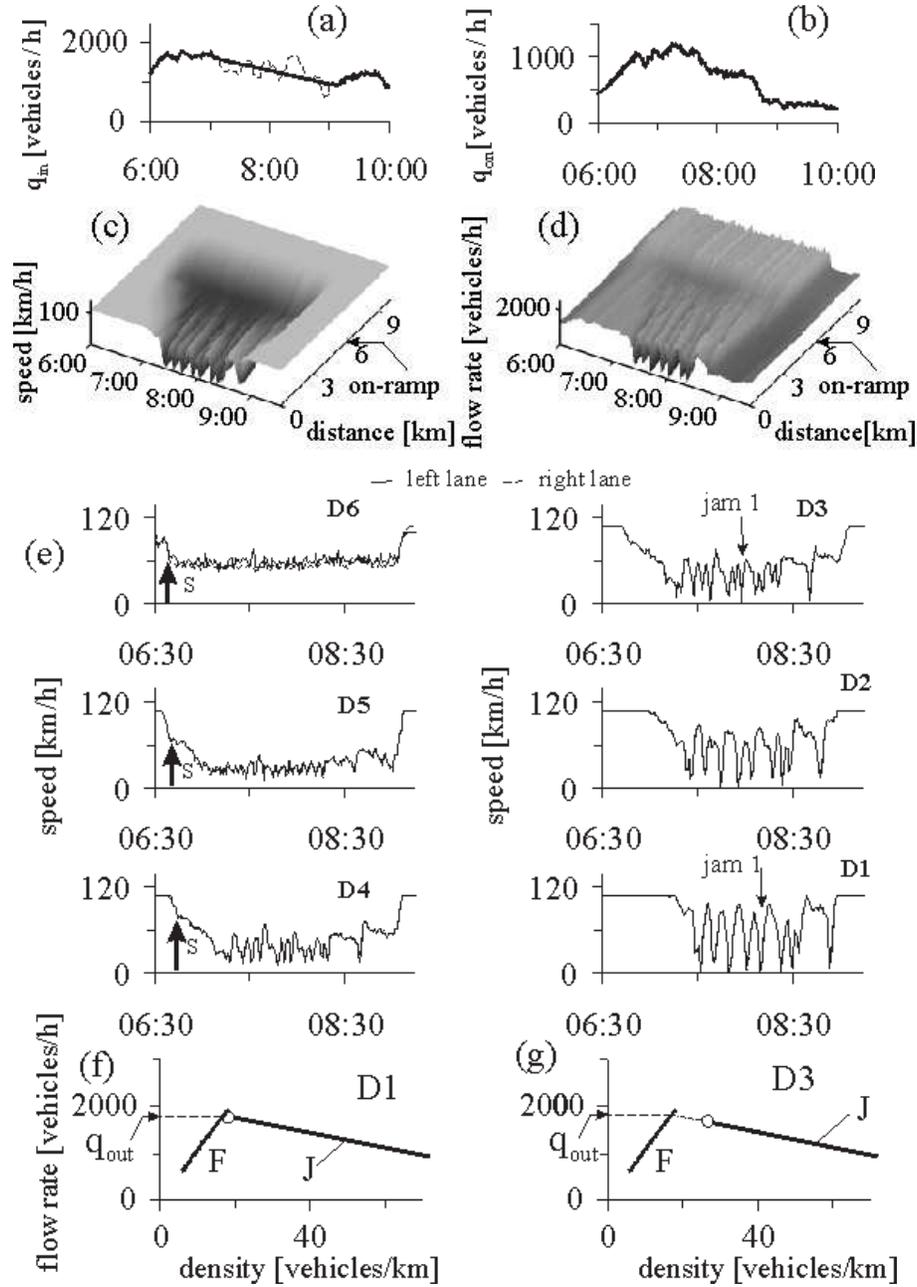}
\caption{Simulation of GP evolution: (a, b) -- Time dependences of flow rates
$q_{\rm in}$ (a) and $q_{\rm on}$ (b). (c, d) --
Speed (c) and flow rate (d) averaged across the road in space and time for spontaneous GP emergence and evolution.
(e) -- Speed at different detectors as a function of time for the GP. 
(f) -- The line $J$ when free flow (left) and synchronized flow (right) is formed in the
wide moving jam outflow for the wide moving jam labeled $\lq\lq$jam 1" within the GP in (e).
$q_{\rm out}$ is the flow rate in free flow formed by
the wide moving jam outflow.
\label{GP} } 
\end{center}
\end{figure*}

\begin{figure*}
\begin{center}
\includegraphics[width=12 cm]{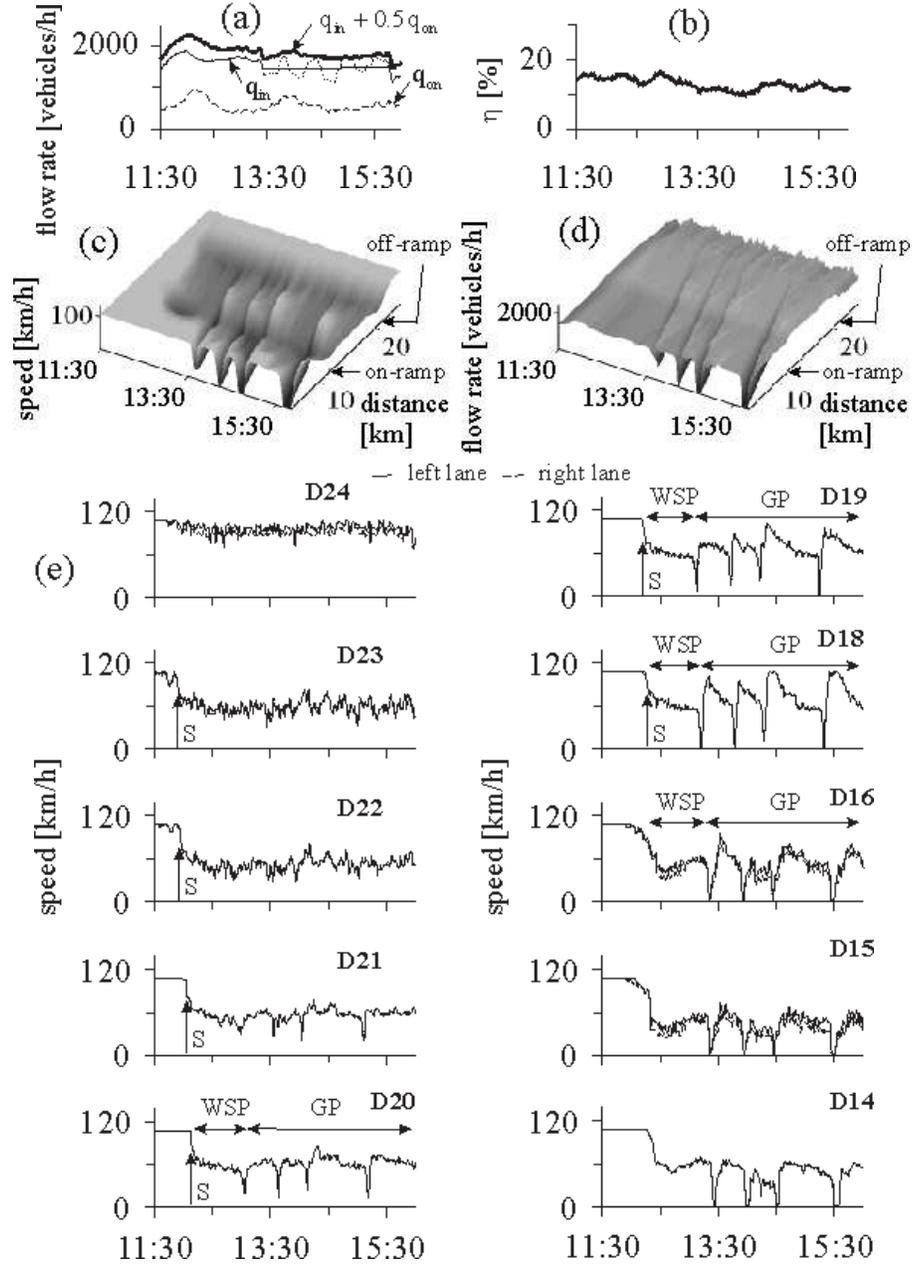}
\caption{Simulations of congested pattern evolution at two adjacent bottlenecks: (a, b) -- Time dependences of flow rates
$q_{\rm in}$, $q_{\rm on}$ (a) and the percentage of vehicles $\eta$ that leave the main road to the off-ramp (b). (c, d) --
Speed (c) and flow rate (d) averaged across the road in space and time.
(e) -- Speed at different detectors as a function of time. 
$v_{\rm free \ off}=$ 65 km/h.
\label{WSP_GP} } 
\end{center}
\end{figure*}

For numerical simulation of WSP emergence at an off-ramp bottleneck and the subsequent WSP transformation 
into an EP and an GP at an on-ramp bottleneck (Fig.~\ref{230301_GP}), the same model as in Sect.~\ref{GP_SC}, however, with 
an off-ramp bottleneck downstream   and an on-ramp
 bottleneck  upstream  
is used.
 At the road upstream boundary, the time-dependence of the flow rate $q_{\rm in}(t)$ associated with empirical
 data measured at the detector D8  (see Fig. 2.2 in~\cite{KernerBook}) has been applied.  During the time interval, when 
  wide moving jams are observed,
  $q_{\rm in}(t)$ is approximated   by a line (Fig.~\ref{WSP_GP} (a)) as in Sect.~\ref{GP_SC}.
 The flow rate to the on-ramp $q_{\rm on}(t)$  is  taken from measurements in the on-ramp lane.
 The empirical flow rate to the off-ramp $q_{\rm off}(t)$ is used to calculate the percentage of vehicles $\eta(t)$
 in the flow rates $q_{\rm in}(t)$ and $q_{\rm on}(t)$, which  leave the main road to the off-ramp
 (Fig.~\ref{WSP_GP} (b)).

In simulations, congested pattern emergence and evolution (Fig.~\ref{WSP_GP})
are qualitatively   the same as those in empirical
 observations (Fig.~\ref{230301_GP}). In particular,  
in accordance with the empirical study (Fig.~\ref{230301_GP}), the following main effects are found:
 
(i) {\it A local first-order F$\rightarrow$S transition occurs at some distance upstream of the off-ramp
bottleneck} (D23, labeled $S$ in Fig.~\ref{WSP_GP} (e)).
A wave of induced F$\rightarrow$S transitions propagates upstream (labeled $S$ in Fig.~\ref{WSP_GP} (e), at D22--D18).
{\it A widening synchronized flow pattern (WSP) is formed} due to these
F$\rightarrow$S transitions. 

(ii) {\it Transformation of the WSP into an expanded pattern (EP).} After the upstream front of the WSP
reaches the upstream bottleneck (on-ramp bottleneck), this front induces an F$\rightarrow$S transition
at the on-ramp bottleneck (D16 in Fig.~\ref{WSP_GP} (e)). Due to subsequent propagation of this synchronized flow front
upstream of the on-ramp bottleneck, the EP occurs in which synchronized flow affects both
downstream and upstream bottlenecks. 

(iii) {\it Intensification of downstream congestion due to upstream congestion.}
After the EP has appeared,
wide moving jams begin to form downstream of the on-ramp bottleneck within synchronized flow of the initial WSP:
The initial WSP transforms into an GP  between the off-ramp and on-ramp bottlenecks.
This intensification of downstream congestion (congestion formation upstream of the off-ramp bottleneck)
due to upstream congestion (congestion  upstream of the on-ramp bottleneck) exhibits the same features as those
in empirical data (Sect.~\ref{WSP_Emp_Sect}).

 (iv) {\it Hysteresis Effect of Pattern Existence.} At $t>$13:30 the flow rates $q_{\rm on}(t)$ and $q_{\rm in}(t)$ 
   become appreciably smaller than at the time of the F$\rightarrow$S transition at the off-ramp bottleneck.
 Nevertheless, due to the hysteresis effect of congested pattern existence the EP persists
 during a very long time affecting both bottlenecks.

\subsubsection{Spatial Dependence of Macroscopic Congested Traffic Pattern Characteristics}
\label{SD_Sect}

\begin{figure*}
\begin{center}
\includegraphics[width=12 cm]{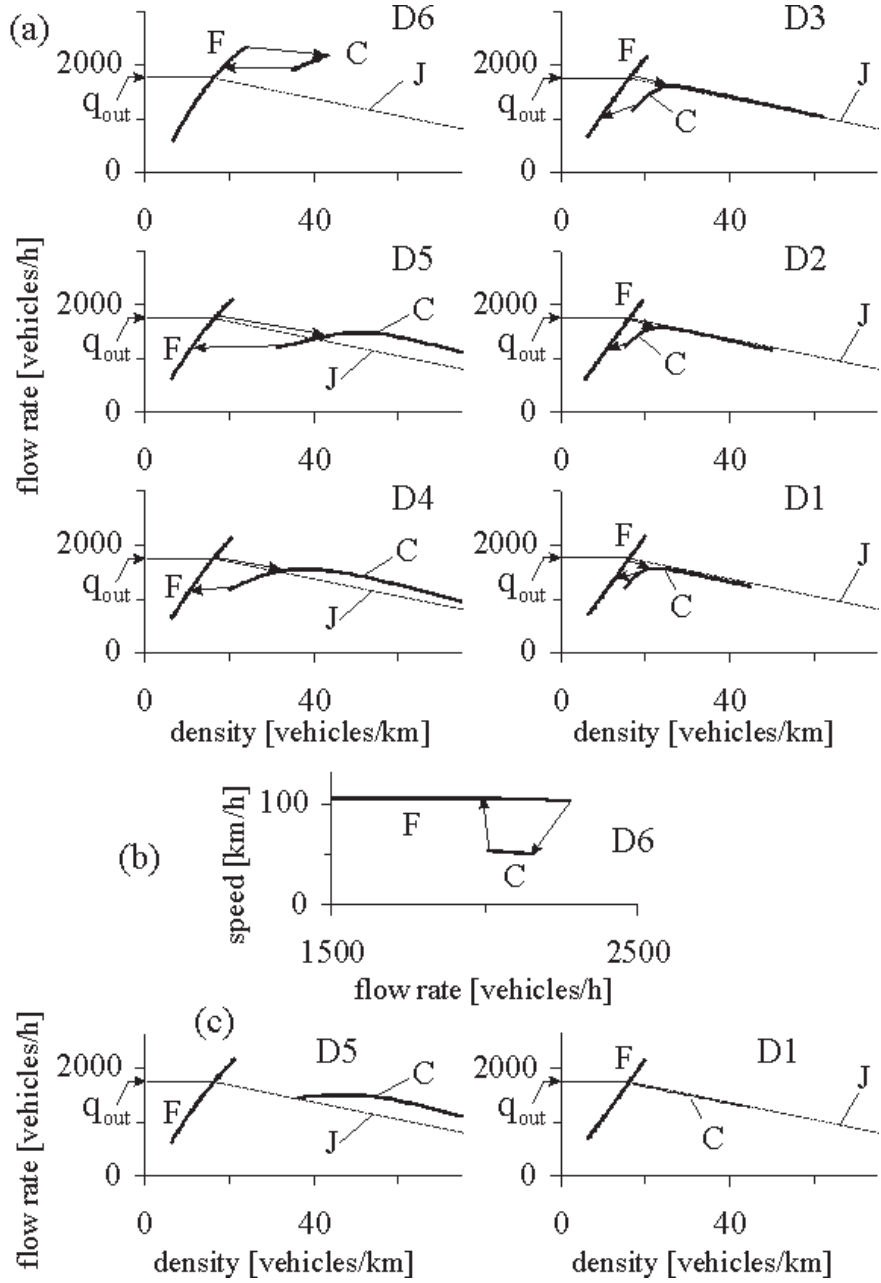}
\caption{Fundamental diagrams and Z-characteristic: (a, b) -- Theoretical spatial 
dependence  of the fundamental diagram (a) and the   Z-characteristic for the
F$\rightarrow$S transition at the bottleneck (D6) (b) for the GP shown
in Fig.~\ref{GP}. (c) -- Theoretical fundamental diagrams for GP under strong congestion.
Arrows in (a, b) show F$\rightarrow$S transitions (from the branch $F$ to the branch $C$)
and S$\rightarrow$F transitions (from the branch $C$ to the branch $F$) at the related detectors. 
\label{GP_FD} } 
\end{center}
\end{figure*}

\begin{figure*}
\begin{center}
\includegraphics[width=12 cm]{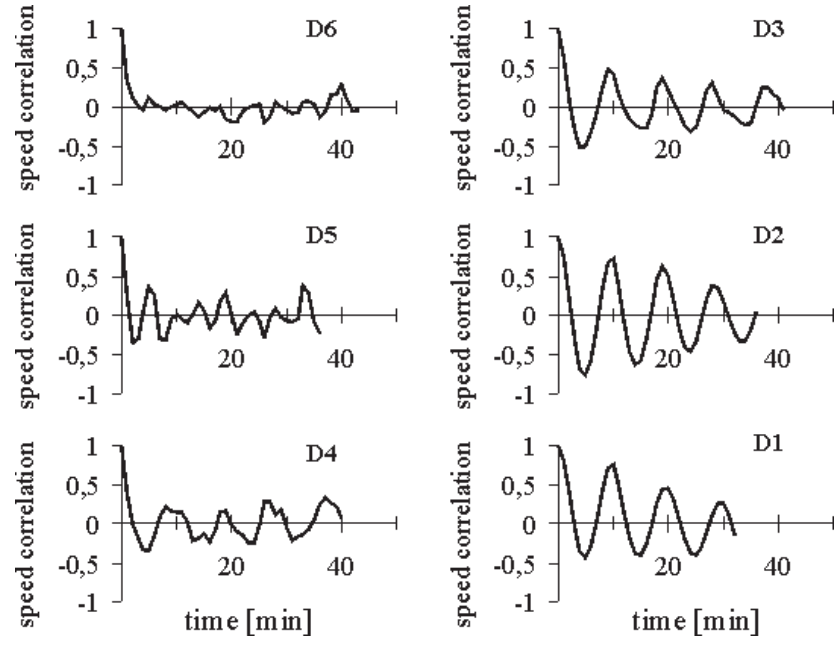}
\caption{Simulated speed correlation functions at different 
detectors for the GP in Fig.~\ref{GP}. 
\label{Corr_S} } 
\end{center}
\end{figure*}

Spatial theoretical dependences of the fundamental diagrams and  speed correlation functions are
qualitatively the same (Figs.~\ref{GP_FD} 
and~\ref{Corr_S}) as those  characteristics in empirical investigations (Fig.~\ref{150496_FD} 
and Fig. 12.8 in~\cite{KernerBook}), respectively.

At the on-ramp bottleneck (D6), the fundamental diagram (Fig.~\ref{GP_FD} (a)) is associated with
the Z-shaped speed--flow rate characteristic (Fig.~\ref{GP_FD} (b)) that is usual for a first-order F$\rightarrow$S
transition~\cite{KernerBook,Z_Hom}; on the branch for synchronized flow
(branch $C$) the flow rate is an increasing density function. In the pinch region of the GP (D5),
the fundamental diagram exhibits a maximum point: At greater densities the flow rate slightly decreases with  density,
whereas at smaller densities the flow rate is a weak increasing density function. At  greater densities associated with the strong
congestion condition, the branch $C$ lies above the line $J$. This is related to growing character of narrow moving jams in the
pinch region. If only the time interval for the strong congestion condition within the pinch region of the GP is considered,
the flow rate is approximately constant: It is only a very weak density function   (D5, Fig.~\ref{GP_FD} (c)).
Upstream  of the pinch region, the greater the distance  a freeway location from the pinch region, the closer 
the branch $C$ to the line $J$ at greater densities. Within the completely formed region of wide moving jams (D2, D1),
the branch $C$ lies on the line $J$ at greater densities. If the time interval in which only wide moving jams are observed
is considered, the branch $C$ lies on the line $J$ for all densities (D1, Fig.~\ref{GP_FD} (c))~\cite{Wide_J}.

The speed correlation function  for the GP (Fig.~\ref{Corr_S})
exhibits  the same features as those in empirical results.

\subsection{Simulations of Microscopic Criterion for Wide Moving Jam}
\label{Interrup_S}

Simulations show that the microscopic criterion for 
  wide moving jam phase  presented in Sect.~\ref{Interruption}
  enable us to distinguish clearly between
the wide moving jam and synchronized flow phases in local microscopic (single vehicle) congested traffic
states.

\begin{figure*}
\begin{center}
\includegraphics[width=12 cm]{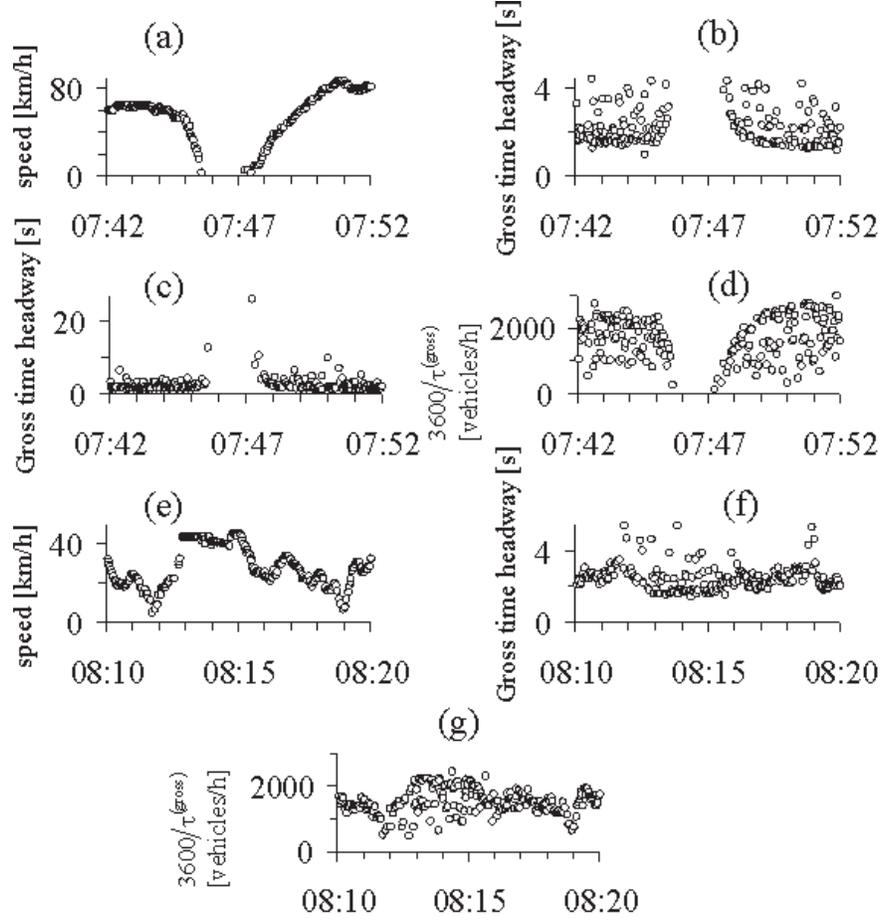}
\caption{Simulations of  microscopic criterion for traffic phase identification
in different local microscopic congested traffic states related to 
different locations within the GP shown in Fig.~\ref{GP} (c--e):
(a--d) -- Single vehicle model data related to the location D2 for speed within a wide moving jam in the left lane (a) and
the associated time distributions of gross time headways $\tau^{\rm (gross)}$ for different scales of the headways (b, c) 
and of the value $3600/\tau^{\rm (gross)}$ (d).
(e--g) -- Single vehicle model data related to the location D4 for speed within a sequence of two narrow moving jams in the left lane (e) and
  the associated time distributions of $\tau^{\rm (gross)}$ (f)
  and of $3600/\tau^{\rm (gross)}$ (g).
\label{Mic_Cr}  }
\end{center}
\end{figure*}

Indeed, in   Fig.~\ref{Mic_Cr} the same dependences and characteristics as those
in empirical results (Fig.~\ref{Iterr_Fig}) are presented for a wide moving jam (Fig.~\ref{Mic_Cr} (a--d))
and two narrow moving jams (Fig.~\ref{Mic_Cr} (e--g)) associated with two different local
microscopic congested traffic states related to two different locations within the GP in Fig.~\ref{GP} (c--e).
A comparison of the empirical  (Fig.~\ref{Iterr_Fig}) with related simulated results (Fig.~\ref{Mic_Cr}) fully confirms
all   conclusions about the traffic phase identification in congested traffic
formulated in Sect.~\ref{Interruption}. In addition, within model wide moving jams there are often gross time headways between vehicles that are
 about  4--7 sec. As found, they are related to moving blanks
 within the jams. This confirms the assumption made in Sect.~\ref{Interruption}
 that such gross time headways within empirical wide moving jams are explained by  moving blanks.

If the microscopic criterion is applied for the moving jams, which propagate through the on-ramp bottleneck in Fig.~\ref{WSP_GP} (e),
then we find that within each of these jams traffic interruption occurs, i.e., the condition
(\ref{GrossDel}) is satisfied. This means that corresponding to the microscopic criterion these moving jams  are wide moving jams. The same conclusion is made from the macroscopic spatiotemporal objective criteria for traffic phases in
congested traffic. Indeed, these jams propagate through the bottleneck while maintaining
the mean jam downstream front velocity. Thus, the microscopic criterion and macroscopic objective spatiotemporal criteria for the traffic phases
lead to the same result by  phase identification in congested traffic. 

\begin{figure*}
\begin{center}
\includegraphics[width=12 cm]{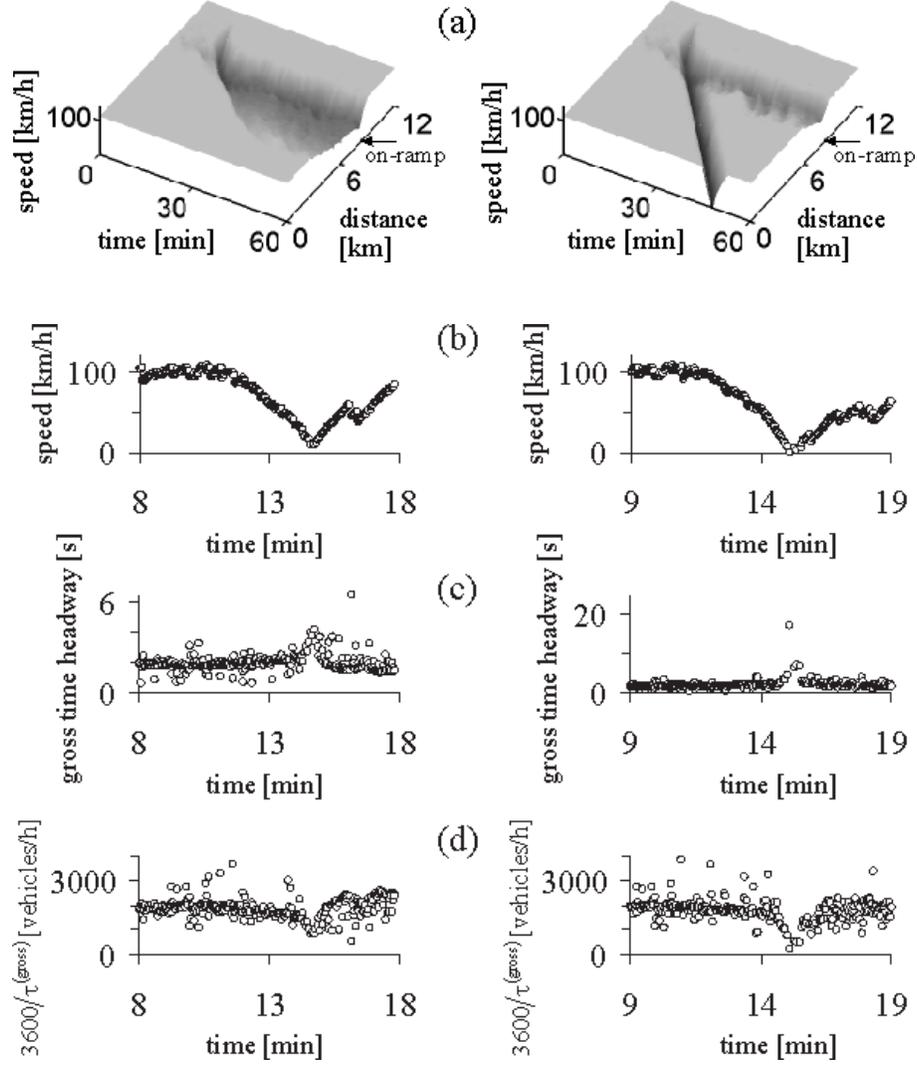}
\caption{Comparison of microscopic criterion with macroscopic spatiotemporal objective  criteria
for the phases in congested traffic. Figures left are related to a narrow moving jam,
 figures right are related to a wide moving jam: (a) -- The catch effect of
 a narrow moving jam at an on-ramp bottleneck with   subsequent LSP formation at the bottleneck (left)
 and wide moving jam propagation through the bottleneck with LSP formation at the bottleneck (right).
 Average vehicle speed (one-minute data) in space and time. (b) --
 Single vehicle model data for vehicle speed.
 (c, d) -- gross time headways (c) and the value $3600/\tau^{\rm (gross})$ (d) related to (b).
$q_{\rm in}=$ 1830, $q_{\rm on}=$ 270, $q_{\rm out}=$ 1810 vehicles/h/lane.
 \label{Mic_Cr_P} } 
\end{center}
\end{figure*}

To prove the latter conclusion, a numerical experiment is performed  additionly (Fig.~\ref{Mic_Cr_P}). In this numerical experiment,
metastable free flow is realized at an on-ramp bottleneck.
Firstly, a narrow moving jam  is induced downstream of the on-ramp bottleneck (Fig.~\ref{Mic_Cr_P} (a), left). 
Indeed, an application of the microscopic criterion
to this moving jam shows (Fig.~\ref{Mic_Cr_P} (b--d), left) that there is no traffic interruption within the jam. 
In this case, we find $\tau^{\rm (gross)}_{\rm max}/\tau^{\rm (ac)}_{\rm del}\approx$ 2
(model time delay $\tau^{\rm (ac)}_{\rm del}\approx$ 1.74 sec). Rather than the jam propagating
through the on-ramp bottleneck, the moving jam is pinned at the bottleneck leading to an F$\rightarrow$S transition with 
subsequent localized SP (LSP) formation upstream of the bottleneck (Fig.~\ref{Mic_Cr_P} (a), left).
Secondly, another moving jam is induced at the same location downstream of the bottleneck (Fig.~\ref{Mic_Cr_P} (a), right). However,
there is flow interruption within this jam (Fig.~\ref{Mic_Cr_P} (b--d), right). Accordingly to the microscopic criterion for 
the phases in congested traffic, this jam is a wide moving jam.
In this case, we find $\tau^{\rm (gross)}_{\rm max}/\tau^{\rm (ac)}_{\rm del}\approx$ 10.  In contrast with former narrow moving jam, this jam
propagates through the bottleneck while maintaning the mean downstream jam front velocity (Fig.~\ref{Mic_Cr_P} (a), right).
As in empirical observations~\cite{KernerBook}, wide moving jam propagation through metastable free flow state at the bottleneck leads
to LSP formation as in the case of the former narrow moving jam. In contrast with the narrow moving jam, synchronized flow
within the LSP has no influence on wide moving jam propagation (Fig.~\ref{Mic_Cr_P} (a), right).  
It is found that to identify a moving jam as a wide moving jam based on the microscopic criterion with certainty
the following numerical relation should be satisfied
$\tau^{\rm (gross)}_{\rm max}/\tau^{\rm (ac)}_{\rm del}>$ 5. 

 The relative large number 5 found for this criterion is associated with fluctuations.
 We have found that in some of different realizations
 made at the same flow rates and initial conditions for jam excitation, when 3 $<\tau^{\rm (gross)}_{\rm max}/\tau^{\rm (ac)}_{\rm del}<$ 5,
 a moving jam is a wide moving jam (it propagates through the bottleneck 
 while maintaining the mean velocity of the jam downstream front), but in the other
 realizations the jam is a narrow moving jam (it is pinned at the bottleneck causing an F$\rightarrow$S transition
 at the bottleneck).
 This is because of fluctuations of vehicle motion within the jam,  in the jam inflow and outflow,
 as well as fluctuations through random vehicle lane changing. In particular, 
 it can turn out that within a moving jam a vehicle changes lane before it passes the detector.
Then  a time headway  measured by the detector  increases. This random lane changing   has obviously no relation to
  flow       interruption within a moving jam. A more detailed study of fluctuations is beyond  the scope of this article.

\subsection{Spatial Dependence of  Single Vehicle Model Characteristics}
\label{Model_Single_Cr}

Here, single vehicle model characteristics associated with the GP at the on-ramp bottleneck
are compared with empirical results. However, the GP in Fig.~\ref{GP} exists about two hours only.
In order to make  more reliable statistical characteristics from   single vehicle model data,
after the flow rates  $q_{\rm on}(t)$ and $q_{\rm in}(t)$ taken from measurements
reach their maximum values, these flow rates do not change in simulations any more.
Then an GP that occurs at the bottleneck does not dissolve over time at all.
The characteristics of this GP are the same as those for the GP in Fig.~\ref{GP} (c--e) before   8:00.

\subsubsection{Model Time Headway Distributions and OV Functions}

\begin{figure*}
\begin{center}
\includegraphics[width=12 cm]{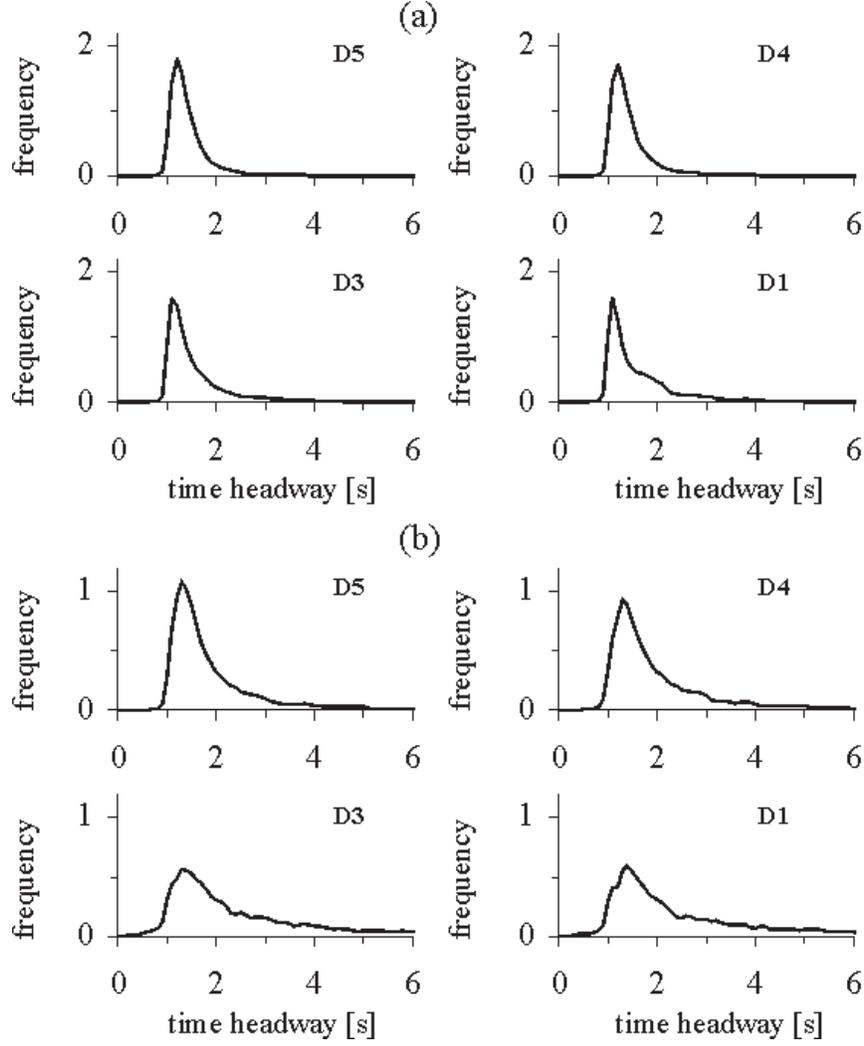}
\caption{Model time headway distributions at different detectors within general patterns: (a) -- Simulations of the model of identical vehicles
related to Fig.~\ref{GP} (e). (b) -- 
Simulations of the model for heterogeneous flow with
various driver and vehicle characteristics of Sect.~20.2 in~\cite{KernerBook} with
30$\%$ of fast, 35$\%$ of slow, and 35$\%$ of long vehicles; $v^{(1)}_{\rm free}=$ 
120 km/h; other parameters for fast vehicles are the same as those for identical vehicles
in Sect.~16.3 of Ref.~\cite{KernerBook};
 $\tau^{\rm(a, \ j)}_{\rm del}(v)=\tau /p^{(j)}_{0}(v)$,
$p^{(j)}_{0}(v)=(a^{(j)} +b^{(j)}\min(1, \ v/10))$, $j=2,3$;
$a^{(2)}=0.42, \ b^{(2)}=0.13$ for slow vehicles ($j=$ 2) and
$a^{(2)}=0.3, \ b^{(2)}=0.18$ for long vehicles ($j=3$).
\label{T_Headways} } 
\end{center}
\end{figure*}

Time headway distributions related to synchronized flow (D5, D4,
Fig.~\ref{T_Headways} (a)) are qualitatively the same as  empirical ones
(Fig.~\ref{TimeHeadways}). 
In the model, time intervals between wide moving jams at D3--D1 (Fig.~\ref{GP} (e)) are considerably shorter than
in the empirical example 1 (Fig.~\ref{Single_Fig} (b)). For  this reason, longer time headways
associated with moving blanks within wide moving jams can be seen
 in the model time headway distributions (D1, Fig.~\ref{T_Headways} (a))
 even without speed separation of time headways made in Fig.~\ref{TimeHeadways} (b).

Quantitative differences between empirical  and model results 
(Figs.~\ref{TimeHeadways} and~\ref{T_Headways} (a))
at small time headways are explained by
 model time step that is equal to 1 sec, i.e., considerably shorter time headways
cannot be observed. 

There are very different vehicles and drivers in real traffic,
specifically aggressive and timid driver behavior are related to shorter and longer 
drivers' time delays, respectively.
In the model of identical vehicles and drivers used above,
chosen driver time delays are more close to aggressive drivers rather than  to timid ones.
This explains why longer time headways do not appear in the model time headway distributions in
Fig.~\ref{T_Headways} (a).
If the microscopic model with various drivers (Sect. 20.2 in~\cite{KernerBook}) is used,  then a quantitative correspondence
with any known empirical result of time headway distributions through the appropriate choice of the percentage of 
slow (timid) drivers and their characteristics is possible (Fig.~\ref{T_Headways} (b)). It is important that in this case the fundamental spatiotemporal features of congested
patterns discussed above do not change, as this has been found in~\cite{KKl2004AA}.

\begin{figure*}
\begin{center}
\includegraphics[width=12 cm]{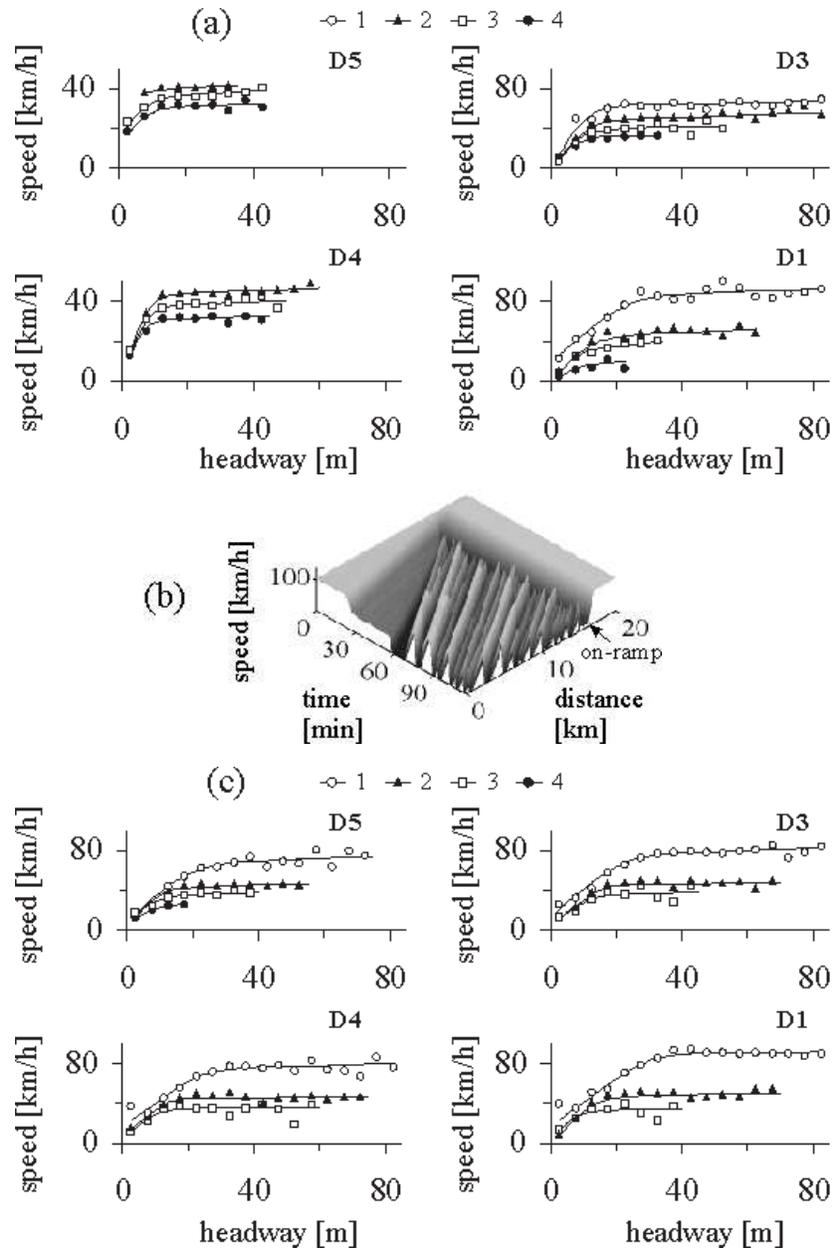}
\caption{Model OV functions at different detectors within general patterns: (a) -- Simulations of the model of identical vehicles
related to Fig.~\ref{GP} (e). (b, c) -- 
Simulations of an GP (b) and associated OV functions (c) 
for the KKW CA model (Sect. 20.2 in~\cite{KernerBook}). Curves 1--4: $\rho=$ 20--30,  $\rho=$ 30--40, $\rho=$ 40--50, and $\rho=$ 50--60
vehicles/km, respectively.
\label{OV_S} } 
\end{center}
\end{figure*}

Model OV functions (Fig.~\ref{OV_S} (a)) show the same characteristics as those for the empirical OV functions (Fig.~\ref{OV_Func}).
This is true for the models with identical vehicles and with various drivers' and vehicles' characteristics.
The same conclusion can be drawn for the KKW CA model (Fig.~\ref{OV_S} (b, c)), in which under the same conditions as those in Fig.~\ref{GP} an GP,
which is qualitatively the same,
 occurs at the on-ramp bottleneck (Fig.~\ref{OV_S} (b))~\cite{KKW_model}.

\subsubsection{Speed Adaptation Functions}

Analyzing  single vehicle data, in Ref.~\cite{Wagner_Lub_2003A}
it has been    found that in congested traffic   a
 distribution $p_{\delta v}(\delta v)$ for vehicle speed difference $\delta v=v_{i+1}-v_{i}$ associated with  two vehicles $i$
and $i+1$,
which are registered at a detector location one after another,  have  a very sharp maximum at $\delta v=0$:
There is an attraction of vehicles to a region with a very
small speed difference in congested traffic.

This behavior has been  explained by the speed adaptation effect~\cite{KKl2004AA,KernerBook}: 
If a vehicle cannot overtake the preceding vehicle, then within a synchronization distance
the vehicle tends  to adapt the vehicle speed to the speed of the preceding vehicle, i.e., the speed
difference $\delta v\rightarrow 0$. The less the probability of overtaken,
the more vehicles due to the speed adaptation effect should move with small speed difference $\delta v$ to each other.
The probability of overtaken is a decreasing function of the speed. 
The decrease of probability of overtaken becomes  appreciable already at higher
densities in free flow. When synchronized flow occurs,
probability of overtaken drops abruptly. 
Specifically,   there is a strong attraction of vehicles in synchronized flow
to a region with small speed difference associated with the speed adaptation effect within the synchronization distance. As a result,
 the function $p_{\delta v}(\delta v)$ has a very sharp maximum in synchronized flow. 
 
\begin{figure*}
\begin{center}
\includegraphics[width=12 cm]{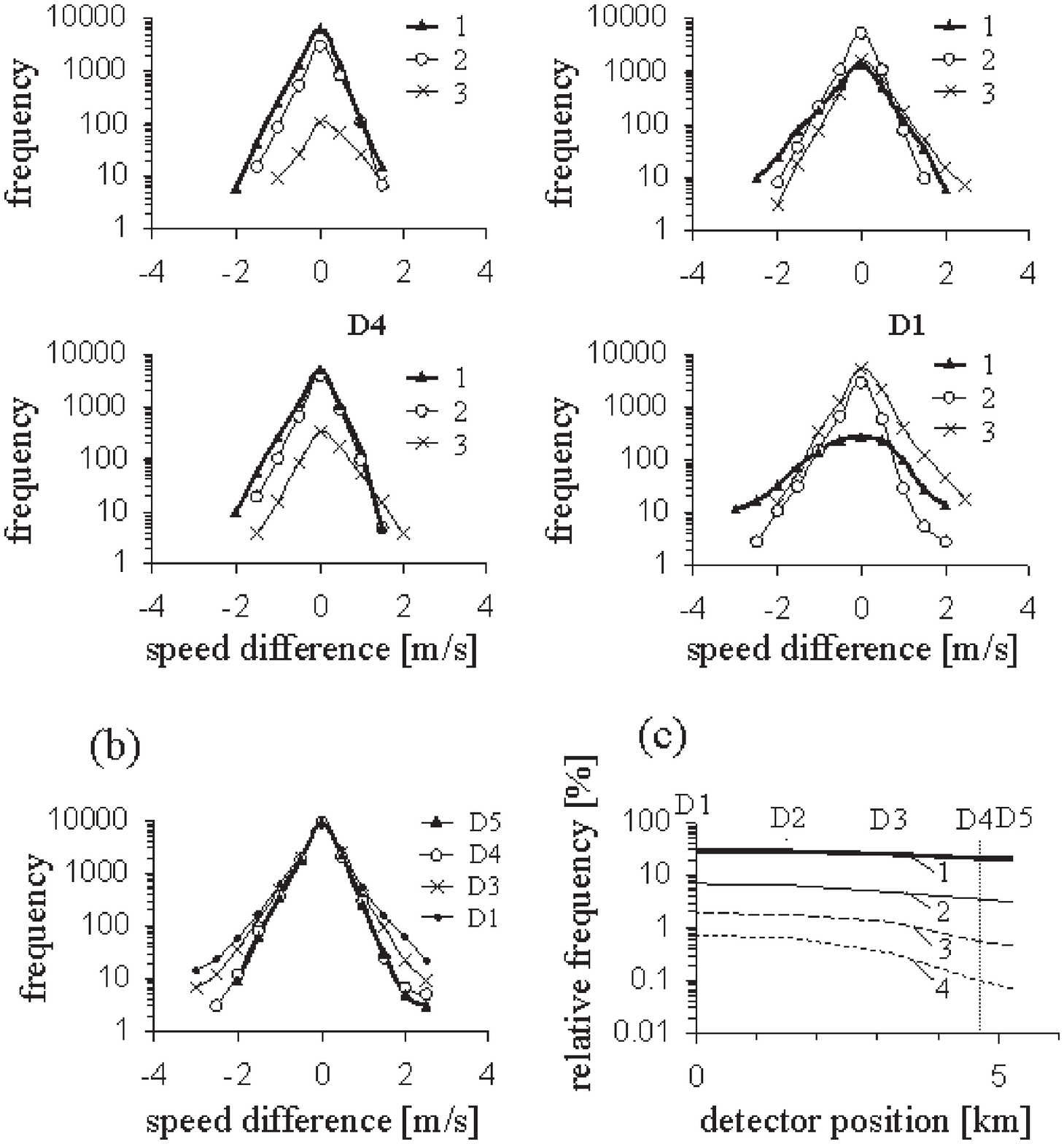}
\caption{Simulations of speed adaptation effect
at different detector locations within
the GP in Fig.~\ref{GP} (e): (a) -- Frequency $p_{\delta v}$ as a function of $\delta v$ for different
ranges of the space headway : curves 1 -- 0--12.5 m, curves 2 -- 12.5--25 m, curves 3 -- $>$ 25 m (densities $\rho>$ 50 vehicles/km,
$30<\rho<50$, and $\rho<$ 30  vehicles/km, respectively). (b) -- 
 Frequency $p_{\delta v}$ as a function of $\delta v$ at different detectors regardless of space headways.
(c) -- Spatial dependence of relative frequency  $p_{\delta v}/p_{0}$ at different given values of the speed difference $\delta v$:
curve 1 -- $\delta v=\pm$ 0.5 m/s, curve 2 -- $\delta v=\pm$ 1 m/s, curve 3 -- $\delta v=\pm$ 1.5 m/s, curve 4 -- $\delta v=\pm$ 2.0 m/s.
$p_{0}=p_{\delta v}\mid_{\delta v=0}$.
\label{Speed_Diff} } 
\end{center}
\end{figure*}

This effect is found in numerical simulations at locations D6 and D5 at which synchronized flow without wide moving jams
 occurs (Fig.~\ref{Speed_Diff} (a)).
However, when wide moving jams appear (D3--D1), then at a given vehicle space headway range
the width of frequency  distribution  $p_{\delta v}(\delta v)$ increases. This means that the 
frequency  distribution  $p_{\delta v}(\delta v)$ in congested traffic is a function of the spatial co-ordinate
and of local dynamics within a congested pattern (Fig.~\ref{Speed_Diff} (b, c)):
 At freeway locations within the pattern in which many wide moving jams propagate (D1, D2, Fig.~\ref{GP} (e)), the 
 distribution $p_{\delta v}(\delta v)$ at a given $\delta v$  becomes greater than at the locations at which synchronized flow without 
 wide moving jams is realized (D5).

\section{Discussion
\label{Discussion}}

Based on empirical and model results presented, the following conclusions can be made:

(i) A   microscopic  criterion 
for the   wide moving jam phase in single vehicle data presented in the article
enables us to identify qualitatively different local microscopic congested traffic states
in both empirical and single vehicle model data measured at a {\it single} freeway location.
   
(ii) A comparison of empirical and simulated  freeway traffic phenomena
   shows that a microscopic three-phase traffic theory of Ref.~\cite{KKl,KKW,KKl2003A,KKl2004AA} can explain   both microscopic and macroscopic
empirical congested pattern features.

(iii) Empirical microscopic (single vehicle) time headway distributions and OV functions found in the article for 
qualitatively different empirical local microscopic congested traffic states can been reproduced
in the microscopic theory for the associated different congested states satisfactorily.

(iv) Time headway distributions, and  distributions for vehicle speed difference
can depend on the spatial co-ordinate considerably. 
However,  the significance of these spatial changes of the traffic characteristics  depends
 on the congested pattern type. In particular, if wide moving jam duration is small in comparison with
 synchronized flow duration, then  time headway distributions and frequency distributions for vehicle speed difference
 do not depend on the freeway location appreciably.

(v)  If 
time-dependencies of traffic demand and drivers' destinations related to macroscopic empirical data
are given at the upstream model boundaries of the main road and an on-ramp, then simulated spatiotemporal
congested patterns emerge,  develop, and dissolve due to self-organization effects in traffic flow in accordance
with empirical observations.

In general,  main features of empirical 
OV functions and time headway distributions are not necessarily   qualitatively different,
even if   local traffic dynamic characteristics within congested patterns are qualitatively different. Therefore,
these and many other empirical traffic flow characteristics (e.g., global and aggregated fundamental diagrams, hysteresis effects, etc.)
 could be considered secondary ones in comparison to spatiotemporal
traffic pattern features (item (1)--(4) in Sect.~\ref{Intr}).

This is because important spatiotemporal features of phase transitions and congested patterns
(item (1)--(4) in Sect.~\ref{Intr}) are {\it lost} in these and many other 
macroscopic and microscopic traffic characteristics, which are widely used as the empirical 
basis for  tests of traffic flow models.
Thus, it is not justified to use these 
macroscopic and microscopic characteristics as the {\it solely empirical basis} for a decision whether a traffic flow model
can describe real traffic flow or not~\cite{NS_TEst}. 

Firstly, a comparison of empirical features of phase transitions in traffic flow and
spatiotemporal congested patterns with associated model solutions for a model of freeway with those freeway bottlenecks, which 
affect the empirical patterns, has to be performed. Then microscopic model characteristics of
single vehicle model data associated with different locations within the patterns
are studied. Finally, these microscopic model characteristics
are compared with    empirical microscopic characteristics of single vehicle data. This data should be related to 
 qualitatively the same empirical   local congested model states as those in the model states.

\begin{thebibliography}{8.}
\bibitem{Aggregated}

Sometimes for better statistical conclusions,
traffic variables obtained
from various days or/and many different locations are aggregated to a data set; this data base of aggregated data is further used for 
 calculation of  traffic flow characteristics.
\bibitem{LongVehicles}

Below the condition (\ref{GrossDel}) is applied for vehicles in the left (passing) freeway line in which person vehicles  move mostly.
When there are many long vehicles in traffic flow, rather than the maximum gross time headway, the maximum time
headway $\tau_{\rm max}$ should be used in the condition (\ref{GrossDel}) for   flow interruption within a moving jam, i.e.,
this condition should be replaced by $\tau_{\rm max}\gg \tau^{\rm (ac)}_{\rm del}$.
This is because at a   low speed of a very long vehicle within a moving jam, the condition (\ref{GrossDel})
   can sometimes be satisfied, if the jam is still a narrow one.
\bibitem{MovingJams}

From a comparison of
a local traffic dynamics measured at D2 (not shown) with the dynamics at D1 in Fig.~\ref{Single_Fig} (b)
(example 1)  has been found that the moving jams propagate upstream with the mean velocity of the downstream
jam fronts $v_{\rm g}\approx -$ 15.8 km/h.
\bibitem{Hysteresis}

It must be noted that during these F$\rightarrow$S$\rightarrow$F hysteresis 
effects traffic flow is not interrupted. This is qualitatively 
 different from F$\rightarrow$J$\rightarrow$F, S$\rightarrow$J$\rightarrow$S, S$\rightarrow$J$\rightarrow$F, and F$\rightarrow$J$\rightarrow$S 
$\lq\lq$hysteresis effects" in the flow--density plane
caused by wide moving jam propagation
through a detector location. 
In the latter cases, there is flow interruption within the jam (Sect.~\ref{Interruption}): 
 The jam inflow $q^{\rm (J)}_{\rm in}$ is independent of the jam outflow $q^{\rm (J)}_{\rm out}$
 ($q^{\rm (J)}_{\rm out}=q_{\rm out}$, if free flow is formed in the   jam outflow).
When $q^{\rm (J)}_{\rm out}<q^{\rm (J)}_{\rm in}$, i.e.,
 the jam width increases over time, the flow rate measured at the detector becomes smaller  after wide moving jam propagation
 through the detector location.
In contrast, when $q^{\rm (J)}_{\rm out}>q^{\rm (J)}_{\rm in}$, i.e.,
the jam width decreases over time,   the flow rate  becomes greater after wide moving jam propagation.
Thus, a  very complex 
$\lq\lq$hysteresis effects" in the flow--density plane are possible. For their understanding,
a spatiotemporal analysis of traffic  pattern emergence and evolution is necessary.
\end{thebibliography}
\end{document}